\documentclass[preprint,showpacs,preprintnumbers,amsmath,amssymb,nofootinbib]
{revtex4}
\usepackage{graphicx}
\usepackage{dcolumn}
\usepackage{bm}
\usepackage{color}
\usepackage{epsfig}
\def\PL #1 #2 #3 {{\it Phys. Lett.} {\bf#1} (#3) #2}
\def\NP #1 #2 #3 {{\it Nucl. Phys.} {\bf#1} (#3) #2}
\def\ZP #1 #2 #3 {{\it Z. Phys.} {\bf#1} (#3) #2}
\def\PRL #1 #2 #3 {{\it Phys. Rev. Lett.} {\bf #1} (#3) #2}
\def\PR #1 #2 #3 {{\it Phys. Rev.} {\bf#1} (#3) #2}
\def\MPL #1 #2 #3 {{\it Mod. Phys. Lett.} {\bf#1} (#3) #2}
\def\RMP #1 #2 #3 {{\it Rev.~Mod. Phys.} {\bf#1} (#3) #2}

\newcommand{\beqn}{\begin{eqnarray}}
\newcommand{\eeqn}{\end{eqnarray}}
\newcommand{\beqns}{\begin{eqnarray*}}
\newcommand{\eeqns}{\end{eqnarray*}}

\newcommand{\beq}{\begin{equation}}
\newcommand{\eeq}{\end{equation}}
\newcommand{\beqa}{\begin{eqnarray}}
\newcommand{\eeqa}{\end{eqnarray}}

\newcommand{\nn}{\nonumber \\}

\begin{document}
\preprint{FERMILAB-PUB-07-419-T}

\title{A Numerical Unitarity Formalism for \\ Evaluating One-Loop Amplitudes}

\author{R.~K.~Ellis}
\email{ellis@fnal.gov}
\author{W~T.~Giele}
\email{giele@fnal.gov}
\affiliation{ Fermilab, Batavia, IL 60510, USA }
\author{Z.~Kunszt}
\email{kunszt@itp.phys.ethz.ch}
\affiliation{ ETH, Zurich, Switzerland }

\date{\today}
\begin{abstract}

Recent progress in unitarity techniques for one-loop scattering
amplitudes makes a numerical implementation of this method
possible. We present a $4$-dimensional unitarity method for
calculating the cut-constructible part of amplitudes and implement the
method in a numerical procedure. Our technique can be applied to any
one-loop scattering amplitude and offers the possibility that one-loop
calculations can be performed in an automatic fashion, as tree-level
amplitudes are currently done. Instead of individual Feynman diagrams,
the ingredients for our one-loop evaluation are tree-level amplitudes,
which are often already known.  To study the practicality of this
method we evaluate the cut-constructible part of the 4, 5
and 6 gluon one-loop amplitudes numerically,
using the analytically known 4, 5 and 6 gluon tree-level 
amplitudes. Comparisons with analytic answers are
performed to ascertain the numerical accuracy of the method.

\end{abstract}
\maketitle
\section{Introduction}

Analytic unitarity techniques in Feynman diagram calculations have
been used for a long
time~\cite{Cutkosky:1960,Diagrammar,vanNeerven:1985xr}. Their use in
the context of gauge theories is even more
powerful~\cite{Bern:1994zx,BDKOneloopInt} and they were successfully
applied to the calculation of one-loop amplitudes of
phenomenologically important 5-leg and 6-leg processes in
QCD~\cite{Zqqgg} (for a recent review see
\cite{Bern:2007dw}).

In gauge theories the
conventional Feynman diagram method produces intermediate
results which are  much more complicated then the final answer.
One evaluates the numerous non gauge-invariant
individual Feynman diagrams by expanding the tensor loop integrals
into form factors. This decomposition generates a large number of terms.
 With a growing number of external particles 
it becomes a forbidding task to simplify the expression
analytically. This forces one to adopt more numerical techniques
(see e.g. ref.~\cite{Ellis:2006ss}), which can be 
computationally intensive due to the large number of terms. 
In addition, the large cancellations between the Feynman diagrams 
can potentially lead to numerical instabilities.

As an alternative, the unitarity cut method uses only on-shell states, 
manipulates gauge invariant amplitudes and has been used
to derive simple answers with simple intermediate steps~\cite{Zqqgg}.
New ideas on  twistors~\cite{WittenTwistor}, 
multipole cuts (generalized unitarity)~\cite{BCFGeneralized},
recursion relations~\cite{BGRecursion,BCFRecursion,BCFW,BDKrecursionOneLoop}, 
algebraic reduction of tensor integrals~\cite{OPP} and 
unitarity in $D$-dimension~\cite{BernMorgan,ABFKM,Mastrolia:2006ki,BFmassive}
have   made the unitarity cut method even more promising.
It appears that ultimately one can find an efficient algorithm
which can be used to calculate the one-loop amplitudes in 
terms of tree-level amplitudes.
The progress is due to three important observations.

First, any one-loop amplitude can be decomposed
in terms of scalar box, triangle and bubble master integrals where
both internal and external particles can be 
massive or off-shell~\cite{Melrose:1965kb,Passarino:1978jh}.
The master integrals\footnote{For a collection of currently 
known one-loop master integrals, see the 
web-site {\tt http://qcdloop.fnal.gov}\,.} 
have to be  calculated in dimensional regularization
and may have infrared (boxes and triangles) or
ultraviolet divergences (bubbles).

The second key observation concerns the application of unitarity
techniques to amplitudes with multiple cuts.
Using unitarity techniques
the  coefficients of the master integrals are determined by
multiple cuts of the amplitude which place the cut internal lines 
on their mass shell.  After cutting, the tree-level  
3-gluon scattering amplitude with all 3 gluons on-shell 
can appear at a vertex of the diagram.
These 3-particle amplitudes are identically zero by momentum conservation.
Therefore coefficients of the associated master integral can not be 
extracted.
This obstacle is removed by the  
observation  that the tree-level helicity amplitudes
can be analytically continued to complex momentum values 
\cite{WittenTwistor,BCFGeneralized}  allowing for
solutions to the unitarity constraints in terms of the complex loop
momentum. All of the relevant tree-level expressions are non-zero and
the appropriate one-loop amplitude is reconstructible from the
tree-level amplitudes.  With this method the coefficients of the
4-point master integrals can be extracted both analytically and
numerically. They are given in terms of the product of 4 tree-level
amplitudes, evaluated with the complex on-shell loop momenta
\cite{BCFGeneralized}.  However the coefficients of the 3- and
2-point master integrals were still difficult to extract because terms
already included in the 4-point contributions had to be subtracted.
It was not clear how to express this subtraction in terms of 
the corresponding tree-level amplitudes.

The third important observation is that there is a systematic
way~\cite{OPP} of
calculating the subtraction terms at the integrand level. By
manipulating the one-loop amplitude before the loop integration is
carried out, the unitarity method is reduced to the algebraic problem
of a multi-pole expansion of a rational function.  Alternatively,
the method of ref.~\cite{OPP} can also be viewed as the calculation 
of the residues of each pole term of the integrand.  
The resulting 4-propagator pole
(i.e. box contribution), 3-propagator pole (triangle contribution) and
2-propagator pole (bubble contribution) naturally decompose the
loop momentum integration vector into a ``physical'' space spanned by
the respective external momenta and the remaining ``trivial'' space
orthogonal to the ``physical'' space.  The so-called spurious terms
(or subtraction terms) of ref.~\cite{OPP} are determined by the most
generally allowed dependence of the residue on the components of the
loop momentum in the ``trivial'' space. By definition, these spurious 
terms vanish upon integration over the loop momentum.
These ideas allow the
extraction of all the coefficients of the master integrals for a given
one-loop amplitude.  A possible algorithm for analytical extraction of
the coefficients within the unitarity method has been worked
out in ref.~\cite{Forde3cut}.

In this paper we will expand on the algebraic method~\cite{OPP} by developing a
numerical scheme. With the numerical method outlined in this paper we
evaluate only the cut-constructible part of the amplitude.  We will
show that the master integral coefficients are calculated in terms of
tree-level amplitudes.  This makes it possible to ``upgrade'' existing
leading order generators to produce the cut-constructible part of the
one-loop amplitudes.  The ``upgrade'' requires allowing two of the
external momenta in the tree-level amplitude to be complex 4-vectors,
while leaving the analytic expression of the tree-level amplitude
unchanged. For example, to evaluate the cut-constructible
part of the 6-gluon amplitude only the analytic 3-, 4-, 5- and 6-gluon 
tree-level amplitudes are needed (expressed in spinor product 
language or any other form). 
Alternatively, one could use an efficient recursive numerical
method to evaluate the tree-level amplitude 
\cite{BGRecursion,BCFRecursion,HELAC}.

Because we implement 4-dimensional unitarity cuts, the so-called
rational part is not generated~\cite{ {Bern:1994zx},{Bern:2007dw}}.
 In principle we could expand our
scheme to $D$-dimensional unitarity cuts, thereby generating the
complete amplitude. However, it is not immediately clear how to do
this while maintaining the requirement of 4-dimensional tree-level building
blocks. Alternative methods exist to determine the rational part,
which in principle can be combined with the numerical method outlined
in this paper to give the complete scattering amplitude. A method for
determining the rational part using on-shell recursion relation has
been successfully developed and used~\cite{Genhel,LoopMHV}. A direct
numerical implementation of this method should be possible and
especially attractive due to the recursive nature of this
method. Direct numerical implementation of the $D$-dimensional
unitarity method~\cite{BernMorgan,ABFKM,Mastrolia:2006ki,BFmassive} is harder but may also
become practical in the near future.  Other methods have been
developed using Feynman diagram expansions. While these methods
are attractive as far as simplicity goes, they re-introduce all the
problems of standard Feynman diagram calculations.  In
ref.~\cite{OPPPhotons} a method for calculating the rational part is
developed partially based on Feynman diagram calculations. In
refs.~\cite{XYZ,BinothPhotons} a method is proposed to use simplifed
Feynman diagram techniques for calculating the rational part.

In section 2 we will derive the formalism, which we then apply in section 3 to calculate the cut-constructible
part of the 4-, 5- and
6-gluon amplitudes at one-loop by only using the tree-level amplitudes for the 4-, 5- and 6-gluon amplitudes.
We will then compare the numerical results to the analytic calculations.

\section{The structure  of the one-loop integrand function}
\begin{figure}[t!]
   \begin{center}
   \leavevmode
   \epsfysize=8.0cm
   \epsffile{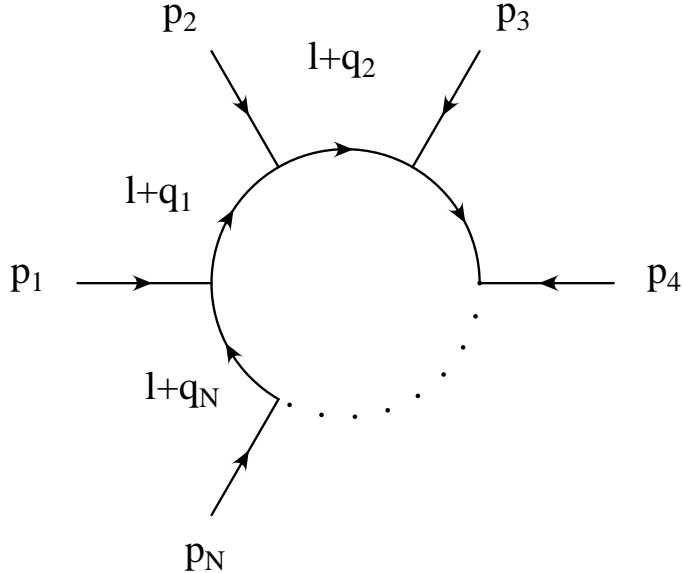}
\end{center}
\caption{The generic $N$-point loop amplitude. }
\label{fig:generic}
\end{figure}

The generic $D$-dimensional $N$-particle one-loop amplitude (fig. \ref{fig:generic})
is given by\footnote{We restrict our discussion to  (color) ordered
external legs. The extension for more general cases is straightforward. } 
\beq
{\cal A}_N(p_1,p_2,\ldots,p_N)=\int [d\,l]\ 
\frac{{\cal N}(p_1,p_2,\ldots,p_N;l)}{d_1d_2\cdots d_N}\ ,
\eeq
where $p_i$ represent the momenta flowing into the amplitude, and $[d \,l]=d^D l$. 
The numerator structure
${\cal N}(p_1,p_2,\ldots,p_N;l)$ is generated by the particle content and is a
function of the inflow momenta and the loop momentum. Since the whole amplitude
has been put on a common denominator, the numerator can also include 
some propagator factors.
The dependence of the amplitude on other quantum numbers has been suppressed. 
The denominator is a product of inverse propagators
\beq\label{parmchoice}
d_i=d_i(l)=(l+q_i)^2-m_i^2=\left(l-q_0+\sum_{j=1}^i p_i\right)^2-m_i^2\ ,
\eeq
where the 4-vector $q_0$ represents the arbitrary parameterization choice of loop momentum. 
The one-loop amplitude in $D=4-2\epsilon$ can be decomposed in the scalar master integral basis giving
\footnote{We drop the finite 6-dimensional 5-point master integral because its coefficient is of ${\cal O}(\epsilon)$~\cite{Bern:1992em}
and therefore it will not contribute to the final answer where we take $\epsilon\rightarrow 0$.}
\beqa\label{MasterDecomp}
{\cal A}_N(p_1,p_2,\ldots,p_N)&=&\sum_{1\leq i_1<i_2<i_3<i_4\leq N} d_{i_1i_2i_3i_4}(p_1,p_2,\ldots,p_N)
 I_{i_1i_2i_3i_4} \nonumber \\
&+&\sum_{1\leq i_1<i_2<i_3\leq N} c_{i_1i_2i_3}(p_1,p_2,\ldots,p_N)
 I_{i_1i_2i_3} \nonumber \\
&+&\sum_{1\leq i_1<i_2\leq N} b_{i_1i_2}(p_1,p_2,\ldots,p_N)
 I_{i_1i_2} \nonumber \\
&+&\sum_{1\leq i_1\leq N} a_{i_1}(p_1,p_2,\ldots,p_N) I_{i_1}\ ,
\eeqa
where the master integrals
 are given by
\beq\label{MasterIntegral}
I_{i_1\cdots i_M}=\int [d\,l]\ \frac{1}{d_{i_1}\cdots d_{i_M}}\ .
\eeq
Analytic expressions for the master integrals with massless internal lines
are reported in ref.~\cite{BDKOneloopInt}.

The maximum number of master integrals is determined by the
dimensionality, $D$, of space-time; for the physical case this gives
up to 4-point master integrals.  The unitarity cut method is based on
the study of the analytic structure of the one-loop amplitude. The
coefficients are rational functions of the kinematical variables and
will in general depend on the dimensional regulator variable 
$\epsilon=(4-D)/2$.  When all the coefficients of the master integrals are
calculated in 4 dimensions we obtain the ``cut-constructible'' part of
the amplitude.  The remaining ``rational part'' is generated by the
omitted ${\cal O}(\epsilon)$ part of the master integral
coefficients~\cite{Bern:2007dw}.

For a numerical procedure we need to recast 
the study of the analytic properties of the unitarity
cut amplitudes into an algebraic algorithm which can be implemented numerically.
In ref.~\cite{OPP} it was proposed that one focus on the integrand of the
one-loop amplitude, 
\beq
{\cal A}_N(p_1,p_2,\ldots,p_N|l)=
\frac{{\cal N}(p_1,p_2,\ldots,p_N;l)}{d_1d_2\cdots d_N} \; .
\eeq
This is a rational function of the loop momentum.  
Any $N$-point tensor integral of rank $M$ ($M\leq N$) with $N\geq 5$ can
be reduced to 4-point tensor integrals of rank $K$ ($K\leq 4$) by application
of Schouten identities. Therefore we can re-express the
rational function in an expansion over 4-, 3-, 2- and 1-propagator
pole terms. The residues of these pole terms contain the master
integral coefficients as well as structures which reside in the subspace
orthogonal to the subspace spanned by the external momenta. 
These spurious terms are important as subtraction terms in the determination of 
lower multiplicity poles. 
The number of spurious structures is 1 for the box, 8 for the triangle,
6 for the bubble, and 4 for the tadpole.
After integration over the loop momenta,
Eq.~(\ref{MasterDecomp}) is recovered. This approach transforms the
analytic unitarity method into the algebraic problem of partial
fractioning a multi-pole rational function. The remaining integrals
after the partial fractioning are guaranteed to be the master integrals
of Eq.~(\ref{MasterIntegral}). This makes a numerical implementation
feasible.

\subsection{The van Neerven-Vermaseren basis}

Consider a set of $R$ inflow momenta, $k_1,\ldots,k_R$ 
in a $D$-dimensional space-time\footnote{The inflow momenta are either 
equal to the external momenta $p_i$, or to sums of external momenta.}.
Taking momentum conservation into account, $\sum_{i=1}^{R} k_i=0$,
the physical space spanned by the momenta $k_i$
has dimension $\min(D,R-1)$. As a consequence, 
for $R\geq D+1$ additional
Schouten identities exist, 
which can be exploited to prove the $D=4$ master integral basis of Eq.~(\ref{MasterDecomp})
\cite{Melrose:1965kb}.
For $R\leq D$, the physical space forms a lower dimensional subspace. We can
define an orthonormal basis, the van Neerven-Vermaseren (NV) basis~\cite{van Neerven:1983vr}, 
which separates the $D$-dimensional space into the $D_P$-dimensional ``physical'' space
and the orthogonal $D_T$-dimensional ``trivial'' space where
\beq
D=D_P+D_T;\ D_P=\min{(D,R-1)};\ D_T=\max{(0,D-R+1)}\ .
\eeq

To define the NV-basis we introduce the generalized Kronecker delta~\cite{VermaserenOldenborgh}
\footnote{
This notation is closely related to the asymmetric Gram determinant 
notation of ref.~\cite{Kajantie},
\[ G\left(\begin{array}{ccc} k_1 & \cdots & k_R \\ q_1 & \cdots & q_R \end{array}\right)=
\delta^{k_1k_2\cdots k_R}_{q_1q_2\cdots q_R}\ .\]}
\beq
\delta^{\mu_1\mu_2\cdots\mu_R}_{\nu_1\nu_2\cdots\nu_R}=\left| \begin{array}{cccc} 
\delta_{\nu_1}^{\mu_1} & \delta_{\nu_2}^{\mu_1} & \dots & \delta_{\nu_R}^{\mu_1} \\
\delta_{\nu_1}^{\mu_2} & \delta_{\nu_2}^{\mu_2} & \dots & \delta_{\nu_R}^{\mu_2} \\
\vdots & \vdots & &\vdots \\ 
\delta_{\nu_1}^{\mu_R} & \delta_{\nu_2}^{\mu_R} & \dots & \delta_{\nu_R}^{\mu_R}
\end{array}\right|\ ,
\eeq
the compact notation
\beq
\delta^{p\mu_2\cdots\mu_R}_{\nu_1q\cdots\nu_R}\equiv
\delta^{\mu_1\mu_2\cdots\mu_R}_{\nu_1\nu_2\cdots\nu_R}k_{\mu_1}q^{\nu_2}\ ,
\eeq
and the $(R-1)$-particle Gram determinant
\beq
\Delta(k_1,k_2,\cdots,k_{R-1})=\delta^{k_1k_2\cdots k_{R-1}}_{k_1k_2\cdots k_{R-1}}\ .
\eeq
Note that for $R\geq D+1$ the generalized Kronecker delta is zero. For the special case $D=R$ we have the factorization of
the Kronecker delta into a product of Levi-Civita tensors:
$\delta^{\mu_1\mu_2\cdots\mu_ R}_{\nu_1\nu_2\cdots\nu_R}=\varepsilon^{\mu_1\mu_2\cdots\mu_R}\varepsilon_{\nu_1\nu_2\cdots\nu_R}$.

Some examples of the generalized Kronecker delta are
\beqa
\delta^{k_1k_2}_{q_1\mu}&=&k_1\cdot q_1\,\delta^{k_2}_{\mu}-{k_1}_{\mu}\delta^{k_2}_{q_1} \nn
&=&k_1\cdot q_1\,{k_2}_{\mu}-k_2\cdot q_1\, {k_1}_{\mu}\nn
\delta^{k_1k_2k_3}_{q_1q_2q_3}&=&k_1\cdot q_1\,\delta^{k_2k_3}_{q_2q_3}
-k_1\cdot q_2\,\delta^{k_2k_3}_{q_1q_3}
+k_1\cdot q_3\,\delta^{k_2k_3}_{q_1q_2} \nn
&=&+k_1\cdot q_1\,(k_2\cdot q_2\,k_3\cdot q_3-k_2\cdot q_3\, k_3\cdot q_2) \nn
& &-k_1\cdot q_2\,(k_2\cdot q_1\,k_3\cdot q_3-k_2\cdot q_3\, k_3\cdot q_1) \nn
& &+k_1\cdot q_3\,(k_2\cdot q_1\,k_3\cdot q_2-k_2\cdot q_2\, k_3\cdot q_1) \,.
\eeqa
We now want to construct the NV-basis for $R$ momenta. 
We define $D_P$ basis vectors
\beq
v_i^\mu (k_1,\ldots,k_{D_P}) \equiv  
 \frac{\delta^{k_1\ldots k_{i-1}\mu k_{i+1}\ldots k_{D_P}}_{k_1\ldots k_{i-1}k_ik_{i+1}\ldots k_{D_P}}}{\Delta (k_1,\ldots,k_{D_P})}\ ,
\label{def:v}
\eeq
with the properties $v_i\cdot k_j=\delta_{ij}$ for $j\leq D_P$. 
When $R\leq D$ we also need to define the projection operator onto the trivial space 
\beq
{w_\mu}^\nu (k_1  \ldots k_{R-1}) \equiv  
\frac{\delta^{k_1\cdots k_{R-1}\nu}_{k_1\ldots k_{R-1}\mu}}{\Delta(k_1,\ldots,k_{R-1})}\ ,
\label{def:w}
\eeq
with the properties ${w_\mu}^\mu=D_T=D+1-R$, $k_i^\mu \, w_{\mu \nu} =0$ and ${w^{\mu}}_{\alpha}w^{\alpha\nu}=w^{\mu\nu}$.
Note that this operator is the metric tensor of the trivial subspace, with the decomposition
\beq\label{metricdecomp}
w^{\mu\nu}=\sum_{i=1}^{D+1-R} n_i^{\mu}n_i^{\nu}\ ,
\eeq
where the $D+1-R$ orthonormal base vectors of the trivial space $n_i$ have the property
$n_i\cdot n_j=\delta_{ij},\ n_i\cdot k_j=n_i\cdot v_j=0$.

The full metric tensor decomposition in the NV-basis is given by
\footnote{By expanding the generalized Kronecker delta functions in the $v_i$ vectors one can show that
$\sum_{i} k_i^\mu v_i ^{\nu}=\sum_{i} k_i^\nu v_i ^{\mu}$.}
\beq\label{fullmetricdecomp}
g^{\mu\nu}=\sum_{i=1}^{D_P} k_i^{\mu}v_i^{\nu}+w^{\mu\nu}
=\sum_{i=1}^{D_P} k_i^{\mu}v_i^{\nu}+\sum_{i=1}^{D_T} n_i^{\mu}n_i^{\nu}\ .
\eeq
For the case $D=R$ the sole basis vector of the 1-dimensional trivial
space is proportional to the Levi-Civita tensor.
For the cases $R<D$ we can explicitly construct the basis vectors fulfilling
all the requirements. 

As an example in the case of $D=4$ and $R=4$ we get
\beqa
v_1^{\mu}(k_1,k_2,k_3)&=&\frac{\delta^{\mu k_2k_3}_{k_1k_2k_3}}{\Delta (k_1,k_2,k_3)};\
v_2^{\mu}(k_1,k_2,k_3)=\frac{\delta^{k_1\mu k_3}_{k_1k_2k_3}}{\Delta (k_1,k_2,k_3)};\
v_3^{\mu}(k_1,k_2,k_3)=\frac{\delta^{k_1k_2\mu }_{k_1k_2k_3}}{\Delta (k_1,k_2,k_3)} \nn
{w_{\mu}}^{\nu}(k_1,k_2,k_3)&=&\frac{\delta^{k_1k_2k_3\nu}_{k_1k_2k_3\mu}}{\Delta(k_1,k_2,k_3)}
={n_1}_{\mu}{n_1}^{\nu}=\frac{\varepsilon_{k_1k_2k_3\mu}\varepsilon^{k_1k_2k_3\nu}}{\Delta(k_1,k_2,k_3)}\ .
\eeqa

We want to decompose the loop momentum into the NV-basis for a graph with denominator factor $d_1,d_2,\ldots,d_R$.
The denominators are as usual given by $d_i=(l+q_i)^2-m_i^2$ and $k_i=q_i-q_{i-1}$. 
By contracting in the loop momentum with the metric tensor given in Eq.~(\ref{fullmetricdecomp})
we get the loop momentum decomposition in the NV-basis 
\beq
l^{\mu}=\sum_{i=1}^{D_P} l\cdot k_i\,v_i^{\nu}+\sum_{i=1}^{D_T} l\cdot n_i\ n_i^{\nu}\ .
\eeq
Using the notation $l\cdot n_i=\alpha_i(l)=\alpha_i$ and the identity
\beq
l\cdot k_i=\frac{1}{2}\left[d_i-d_{i-1}-\left(q_i^2-m_i^2\right)+\left(q_{i-1}^2-m_{i-1}^2\right)\right]\ ,
\eeq
we find
\beq
l^\mu = V_R^{\mu}+\sum_{i=1}^{D_P} \frac{1}{2}(d_i-d_{i-1})\, v_i^{\mu}+\sum_{i=1}^{D_T}\alpha_i\, n_i^\mu\ ,
\label{eq:l3} 
\eeq
where $d_0=d_R$, $m_0=m_R$ and
\beq
V_R^\mu=-\frac{1}{2}\sum_{i=1}^{D_P} \Big((q_i^2-m_i^2)-(q_{i-1}^2-m_{i-1}^2)\Big)\, v_i^\mu\ .
\eeq
In the case that $R\geq D+1$ the decomposition of the loop momentum into the NV-basis 
implicitly proves Eq.~(\ref{MasterDecomp}).
Also, when $R\leq D$ it allows us to include the unitarity constraints without resorting to
the explicit 4-dimensional spinor formalism used in analytic calculations. By avoiding the 4-dimensional
spinor formalism, the formulation is also valid for massive internal particles (where the mass can be real or complex valued).

For example, in the case of a 4-dimensional pentagon, ($D=4$ and $R=5$) we get
\beqa
l^\mu&=&V_5^{\mu}+\frac{1}{2}(d_1-d_5)\, v_1^{\mu}+\frac{1}{2}(d_2-d_1)\, v_2^{\mu}+\frac{1}{2}(d_3-d_2)\, v_3^{\mu}
+\frac{1}{2}(d_4-d_3)\, v_4^{\mu} \nn
V_5^{\mu}&=&-\frac{1}{2}(q_1^2-q_5^2-m_1^2+m_5^2)\, v_1^{\mu}-\frac{1}{2}(q_2^2-q_1^2-m_2^2+m_1^2)\, v_2^{\mu}\nn
&&-\frac{1}{2}(q_3^2-q_2^2-m_3^2+m_2^2)\, v_3^{\mu}-\frac{1}{2}(q_4^2-q_3^2-m_4^2+m_3^2)\, v_4^{\mu} \; .
\eeqa
Similarly for a 4-dimensional triangle ($D=4$ and $R=3$) we get
\beqa
l^\mu&=&V_3^{\mu}+\frac{1}{2}(d_1-d_3)\, v_1^{\mu}+\frac{1}{2}(d_2-d_1)\, v_2^{\mu}
+\alpha_1\, n_1^{\mu}+\alpha_2\, n_2^{\mu} \nn
V_3^{\mu}&=&-\frac{1}{2}(q_1^2-q_3^2-m_1^2+m_3^2)\, v_1^{\mu}-\frac{1}{2}(q_2^2-q_1^2-m_2^2+m_1^2)\, v_2^{\mu} \; .
\eeqa
Thus we see that the same basis decomposition is used for the tensor reductions in the case $R\geq D+1$ and
solving the unitarity constraint in the case that $R\leq D$.

\subsection{Partial fractioning of the integrand}

For the remainder of the paper we restrict ourselves to a 4-dimensional space.
Given the master integral decomposition of Eq.~(\ref{MasterDecomp}) we can partial fraction the integrand
of any 4-dimensional $N$-particle amplitude as
\beq
{\cal A}_N(l)
=\!\!\!\!\!\sum_{1\leq i_1<i_2<i_3<i_4\leq N} \frac{\overline{d}_{i_1i_2i_3i_4}(l)}{d_{i_1}d_{i_2}d_{i_3}d_{i_4}}
+\!\!\!\!\sum_{1\leq i_1<i_2<i_3\leq N} \frac{\overline{c}_{i_1i_2i_3}(l)}{d_{i_1}d_{i_2}d_{i_3}}
+\!\!\!\sum_{1\leq i_1<i_2\leq N} \frac{\overline{b}_{i_1i_2}(l)}{d_{i_1}d_{i_2}}
+\!\!\sum_{1\leq i_1\leq N} \frac{\overline{a}_{i_1}(l)}{d_{i_1}}\,.\nn
\eeq

To calculate the numerator factors, we will calculate the residues by taking 
the inverse propagators equal to zero.
The residue has to be taken
by constructing the loop momentum $l_{ij\cdots k}$ such that 
$d_i(l_{ij\cdots k})=d_j(l_{ij\cdots k})=\cdots=d_k(l_{ij\cdots k})=0$. 
Then the residue of a function $F(l)$ is given by
\beq
\mbox{Res}_{ij\cdots k} \left[F(l)\right] \equiv
\left.\Big(d_i(l)d_j(l)\cdots d_k(l) F\left(l\right)\Big)\right\rfloor_{l=l_{ij\cdots k}}\ .
\eeq
The specific residues are now given by
\beqa
\overline{d}_{ijkl}(l)&=&\mbox{Res}_{ijkl}\Big({\cal A}_N(l)\Big)\nn
\overline{c}_{ijk}(l)&=&\mbox{Res}_{ijk}
\left({\cal A}_N(l)-\sum_{l\neq i,j,k}\frac{\overline{d}_{ijkl}(l)}{d_id_jd_kd_l}\right) \nn
\overline{b}_{ij}(l)&=&\mbox{Res}_{ij}
\left({\cal A}_N(l)-\sum_{k\neq i,j}\frac{\overline{c}_{ijk}(l)}{d_id_jd_k}
-\frac{1}{2!}\sum_{k,l\neq i,j}\frac{\overline{d}_{ijkl}(l)}{d_id_jd_kd_l}\right) \nn
\overline{a}_{i}(l)&=&\mbox{Res}_i
\left({\cal A}_N(l)-\sum_{j\neq i}\frac{\overline{b}_{ij}(l)}{d_id_j}
-\frac{1}{2!}\sum_{j,k\neq i}\frac{\overline{c}_{ijk}(l)}{d_id_jd_k}
-\frac{1}{3!}\sum_{j,k,l\neq i}\frac{\overline{d}_{ijkl}(l)}{d_id_jd_kd_l}\right)\ . \nn
\eeqa
Note that the coefficients are defined to be symmetric in the propagator indices
(e.g. $\overline{c}_{123}=\overline{c}_{213}=\overline{c}_{312}$)
and coefficients with repeated indices are to be set to zero (e.g. $d_{1336}=c_{112}=0$).

As an example, some residues of a 5-particle amplitude are given by
\beqa
\overline{d}_{1245}&=&\mbox{Res}_{1245}\Big({\cal A}_N(l)\Big)\nn
\overline{c}_{235}&=& \mbox{Res}_{235}\left({\cal A}_N(l)
-\frac{\overline{d}_{1235}(l)}{d_1d_2d_3d_5}-\frac{\overline{d}_{2345}(l)}{d_2d_3d_4d_5}\right)\nn
\overline{b}_{14}&=&\mbox{Res}_{14}\left({\cal A}_N(l)
-\frac{\overline{c}_{124}(l)}{d_1d_2d_4}-\frac{\overline{c}_{134}(l)}{d_1d_3d_4}-\frac{\overline{c}_{145}(l)}{d_1d_4d_5}
-\frac{\overline{d}_{1234}(l)}{d_1d_2d_3d_4}-\frac{\overline{d}_{1245}(l)}{d_1d_2d_4d_5}-\frac{\overline{d}_{1345}(l)}{d_1d_3d_4d_5}\right)\ .\nn
\eeqa

In the following sub-sections we will explicitly construct the residue functions
using only tree-level amplitudes. This construction is well-suited for numerical implementation.\\

\subsection{Constructing the box residue}

\begin{figure}[t!]
   \begin{center}
   \leavevmode
   \epsfysize=8.0cm
   \centerline{\epsfig{figure=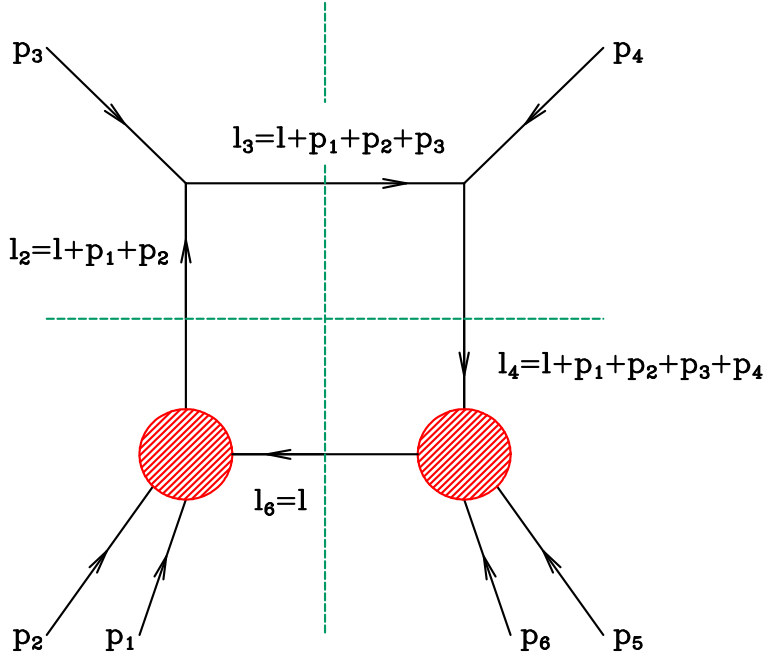,angle=-90,width=10cm}}
\end{center}
\caption{The factorization of the 6-gluon amplitude for the calculation of the $\overline{d}_{2346}(l)$ residue
with the loop momentum parametrization choice $q_0=0$. }
\label{fig:fig2}
\end{figure}

To calculate the box coefficients we choose the loop momentum $l_{ijkl}$ 
such that four inverse propagators are equal to zero, 
\beq\label{Resdef4}
\overline{d}_{ijkl}(l_{ijkl})=\mbox{Res}_{ijkl}\Big({\cal A}_N(l)\Big)\ .
\eeq
We will drop the subscripts on the loop momentum in the following.
Because we have to solve the unitarity constraints explicitly, we
have to choose a specific parameterization, $q_0$, in Eq.~(\ref{parmchoice}).
Using the NV-basis of the four inflow momenta for the box 
and using the fact that $d_i=d_j=d_k=d_l=0$ 
we can use Eq.~(\ref{eq:l3}) to decompose the loop momentum as
\beq\label{Ldef4}
l^\mu=V_4^{\mu}+\alpha_1\, n_1^{\mu}\ .
\eeq
Choosing  for the parameterization $q_0=\sum_{j=1}^l p_j$ (as usual the index $i$ is understood to be modulo $N$) such that $q_l=0$
we have 
\beq\label{Vdef4}
V_4^{\mu}=
-\frac{1}{2}(q_{i}^2-m_{i}^2+m_{l}^2)\, v_1^{\mu}
-\frac{1}{2}(q_{j}^2-q_{i}^2-m_{j}^2+m_{i}^2)\, v_2^{\mu}
-\frac{1}{2}(q_{k}^2-q_{j}^2-m_{k}^2+m_{j}^2)\, v_3^{\mu}\ ,
\eeq
where
\beq\label{videf4}
v_1^{\mu}=\frac{\delta^{\mu k_2k_3}_{k_1k_2k_3}}{\Delta(k_1,k_2,k_3)};\
v_2^{\mu}=\frac{\delta^{k_1\mu k_3}_{k_1k_2k_3}}{\Delta(k_1,k_2,k_3)};\
v_3^{\mu}=\frac{\delta^{k_1k_2\mu }_{k_1k_2k_3}}{\Delta(k_1,k_2,k_3)};\
n_1^{\mu}=\frac{\varepsilon^{\mu k_1k_2k_3}}{\sqrt{\Delta(k_1,k_2,k_3)}}\ ,
\eeq
and
\beq
k_1=q_i;\ k_2=q_j-q_i;\ k_3=q_k-q_j;\ \Delta(k_1,k_2,k_3)=\delta^{k_1k_2k_3}_{k_1k_2k_3}\ .
\eeq
The variable $\alpha_1$ will be determined such that the unitarity condition $d_i=d_j=d_k=d_l=0$ is fulfilled\footnote{
In fact $\alpha_1=\alpha_1(l)$, by varying the loop momentum (allowing for complex values) we can set
$\alpha_1$ to the required complex value. This dependence is implicitly assumed, we will treat $\alpha_1$ as
a ``free variable''.}.
Imposing the constraint $d_n=0$ for $n=\{i,j,k,l\}$ using Eq.~(\ref{Ldef4}) 
\beqa
d_n=0&=&(l+q_n)^2-m_n^2\nn
&=&(V_4+\alpha_1 n_1+q_n)^2-m_n^2\nn
&=&V_4^2+\alpha_1^2+2V_4\cdot q_n+q_n^2-m_n^2\ ,
\eeqa
and using the identity 
\beq
V_4\cdot q_n = -\frac{1}{2}(q_n^2-m_n^2+m_l^2)\ ,
\eeq
so that
\beq 
\alpha_1^2 = -( V_4^2-m_l^2)\ ,
\eeq
we find two complex solutions
\beq \label{boxtwocomplexsolns}
l_{\pm}^{\mu}=V_4^{\mu}\pm i\,\sqrt{V_4^2-m_l^2}\times n_1^{\mu}\ ,
\eeq
which are easily numerically implemented.
We note that because $d_i=d_j=d_k=d_l=0$, we have 
$(l_n^{\pm})^2=m_n^2$ for $n=i,j,k,l$ where $l_n^{\pm}=l^{\pm}_{ijkl}+q_n=l^{\pm}_{ijkl}+\sum_{j=l+1}^n p_i$.
In other words the four propagators are on-shell and the amplitude
factorizes for a given intermediate state into 
4 tree-level amplitudes ${\cal M}^{(0)}$. For the residue of the amplitude in Eq.~(\ref{Resdef4}) 
we find (all indices are assumed modulo $N$, i.e. $i=n+N=n$)
\beqa
\mbox{Res}_{ijkl}\Big({\cal A}_N(l^\pm)\Big)&=&
{\cal M}^{(0)}(l_i^\pm;p_{i+1},\ldots,p_{j};-l_j^\pm)\times
{\cal M}^{(0)}(l_j^\pm;p_{j+1},\ldots,p_{k};-l_k^\pm)\nn &\times&
{\cal M}^{(0)}(l_k^\pm;p_{k+1},\ldots,p_{l};-l_l^\pm)\times
{\cal M}^{(0)}(l_l^\pm;p_{l+1},\ldots,p_{i};-l_i^\pm)\ ,
\eeqa
where the loop momenta $l_n^{\mu}$ are complex on-shell momenta and 
there is an implicit sum over all states of the cut lines (such as e.g. particle type, color, helicity).
For example, the residue of the amplitude for the pure 6-gluon ordered amplitude with $d_6=d_2=d_3=d_4=0$ factorizes into (see fig.~\ref{fig:fig2})
\beqa
\mbox{Res}_{2346}\Big({\cal A}_6(l^\pm)\Big)&=&
{\cal M}_4^{(0)}(l_6^\pm;p_1,p_2;-l_2^\pm)\times{\cal M}_3^{(0)}(l_2^\pm;p_3;-l_3^\pm)\nn
&\times&{\cal M}_3^{(0)}(l_3^\pm;p_4;-l_4^\pm)\times
{\cal M}_4^{(0)}(l_4^\pm;p_5,p_6;-l_6^\pm)\ ,
\eeqa
where the implicit sum over the two helicity states of the four cut gluons is assumed. 
The tree-level 3-gluon amplitudes,
${\cal M}_3^{(0)}$, are non-zero because the two cut gluons have complex momenta \cite{BCFGeneralized}.

Any remaining dependence of the residue $\overline{d}_{ijkl}$ on the 
loop momentum enters through its component in the trivial space,
\beq
\overline{d}_{ijkl}(l) \equiv \overline{d}_{ijkl}(n_1\cdot l)\ .
\eeq
The number of powers of the loop momentum $l$ in the numerator structure 
is called the rank of the integral. 
After integration we find using Eq.~(\ref{metricdecomp}) 
that $(n_1\cdot l)^2\sim n_1^2=1$. Thus rank one is the maximum 
rank of a spurious term (which by definition vanishes upon integration over $l$).
Hence the most general form of the residue is 
\beq
\overline{d}_{ijkl}(l)=d_{ijkl}+\tilde{d}_{ijkl}\,l\cdot n_1\ .
\eeq
Using the two solutions of the unitarity constraint, 
Eq.~(\ref{boxtwocomplexsolns}),
we now can determine the two coefficients of the residue
\beqa\label{Cdef4}
d_{ijkl}&=&\frac{\mbox{Res}_{ijkl}\Big({\cal A}_N(l^+)\Big) +\mbox{Res}_{ijkl}\Big({\cal A}_N(l^-)\Big)}{2}\nn 
\tilde{d}_{ijkl}&=&\frac{\mbox{Res}_{ijkl}\Big({\cal A}_N(l^+)\Big) - \mbox{Res}_{ijkl}\Big({\cal A}_N(l^-)\Big)}{2i\sqrt{V_4^2-m_l^2}}\ .
\eeqa

With the above prescription it is now easy to determine the  spurious term 
for any value of the loop momentum. Finally we note that the integration over the term
\beq
\int [d\,l]\ \frac{\overline{d}_{ijkl}(l)}{d_id_jd_kd_l}=
\int [d\,l]\ \frac{d_{ijkl}+\tilde{d}_{ijkl}\,n_1\cdot l}{d_id_jd_kd_l}=
d_{ijkl}\int [d\,l]\ \frac{1}{d_id_jd_kd_l}=d_{ijkl}I_{ijkl}\ ,
\eeq
is now trivially done, giving us the coefficient of the box times the box master integral.

\subsection{Construction of the triangle residue}

To calculate the triangle coefficients we need to put three propagators on-shell. Care has
to be taken to remove the box contributions by explicit subtraction. Thus, the triangle
coefficient is given by
\beq\label{trianglecoeff}
\overline{c}_{ijk}(l)=\mbox{Res}_{ijk}
\left({\cal A}_N(l)-\sum_{l\neq i,j,k}\frac{\overline{d}_{ijkl}(l)}{d_id_jd_l}\right)\ .
\eeq
Decomposing the loop momentum in the NV-basis of the three inflow momenta of the triangle with $d_i=d_j=d_k=0$
(choosing $q_k=0$)  gives us
according to Eq.~(\ref{eq:l3})
\beq\label{Ldef3}
l^\mu=V_3^\mu+\alpha_1 n_1^\mu+\alpha_2 n_2^\mu\ ,
\eeq
with
\beq\label{Vdef3}
V_3^{\mu}=
-\frac{1}{2}(q_{i}^2-m_{i}^2+m_{k}^2)\, v_1^{\mu}
-\frac{1}{2}(q_{j}^2-q_{i}^2-m_{j}^2+m_{i}^2)\, v_2^{\mu}\ ,
\eeq
where
\beq\label{videf3}
v_1^{\mu}=\frac{\delta^{\mu k_2}_{k_1k_2}}{\Delta(k_1,k_2)};\
v_2^{\mu}=\frac{\delta^{k_1\mu }_{k_1k_2}}{\Delta(k_1,k_2)}\ ,
\eeq
and
\beq
k_1=q_i;\ k_2=q_j-q_i;\ \Delta(k_1,k_2)=\delta^{k_1k_2}_{k_1k_2}\ .
\eeq
The base vectors of the trivial space $\{n_1,n_2\}$ have to be explicitly constructed
using the constraints
\beq\label{Ndef3}
n_i\cdot n_j=\delta_{ij};\ n_i\cdot k_j=0;\ w^{\mu\nu}(k_1,k_2,k_3)=n_1^\mu n_1^\nu+n_2^\mu n_2^\nu\ .
\eeq
The unitarity constraints ($d_i=d_j=d_k=0$) give an 
infinite set of solutions
\footnote{For massless internal lines the parameterization of ref.~\cite{Forde3cut}
is obtained by taking 
$\alpha_1=\kappa\times(a+i\,b)$ and $\alpha_2= \kappa\times(a-i\,b)$ where $ \kappa^2= -V_3^2$,
$a=\frac{1}{2}(t - 1/t)$ and $b=\frac{1}{2}(t + 1/t)$. By taking the $t\rightarrow\infty$ limit one gets
the coefficient of the triangle master integrals.}
\beq
l_{\alpha_1\alpha_2}^{\mu}=V_3^\mu+\alpha_1\,n_1^\mu+\alpha_2\,n_2^\mu;\ \alpha_1^2+\alpha_2^2=-(V_3^2-m_k^2) \; .
\eeq
\begin{figure}[t!]
   \begin{center}
   \leavevmode
   \epsfysize=8.0cm
   \centerline{\epsfig{figure=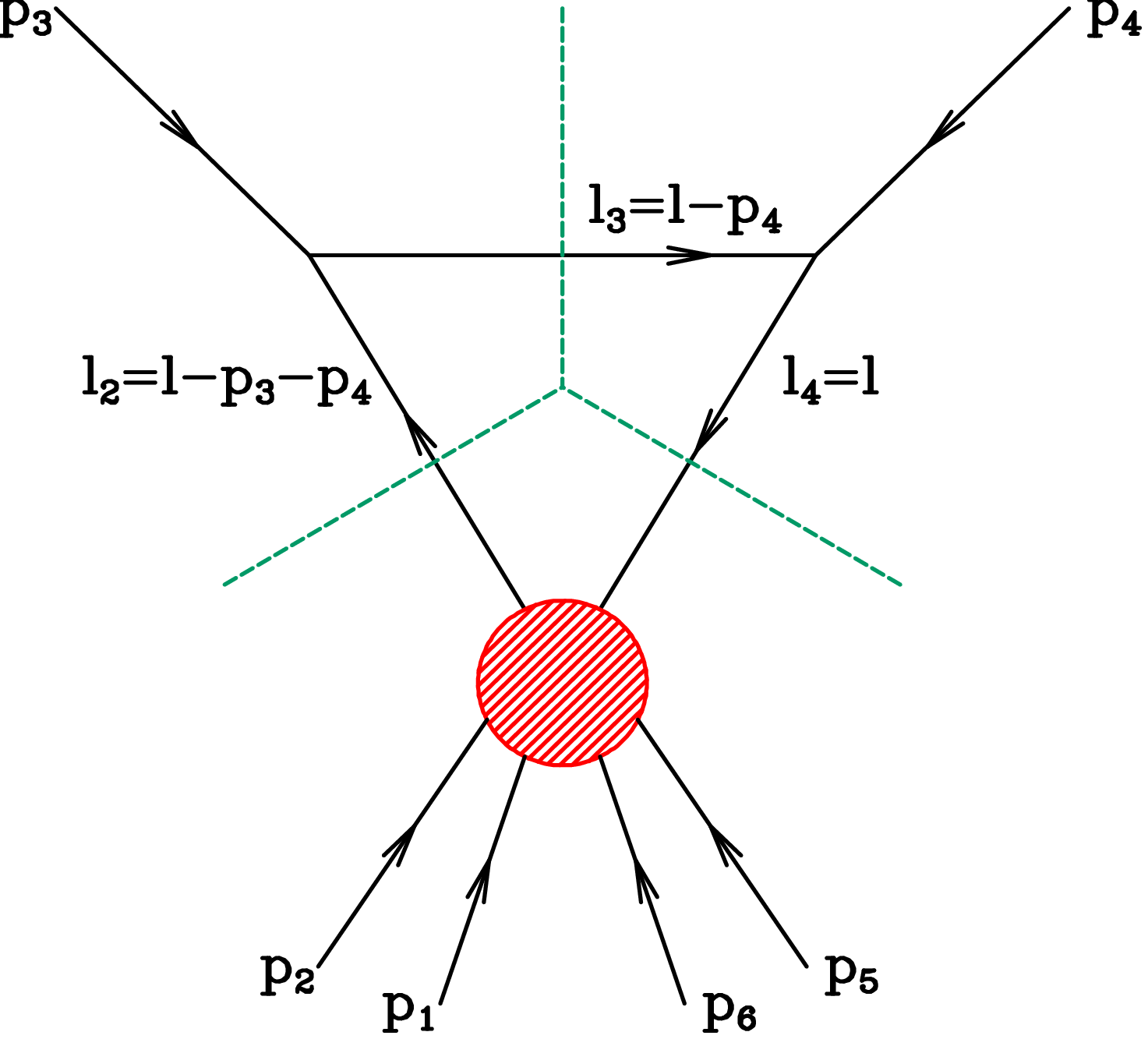,angle=-90,width=10cm}}
\end{center}
\caption{The factorization of the 6-gluon ordered amplitude for the calculation of the $\overline{c}_{234}(l)$ residue
with the loop momentum parametrization choice $q_0=-p_5-p_6=p_1+p_2+p_3+p_4$. }
\label{fig:fig3}
\end{figure}
In this case the residue of the amplitude factorizes into three tree-level amplitudes
\beqa
\mbox{Res}_{ijk}\Big({\cal A}_N(l^{\alpha_1\alpha_2})\Big)&=&
{\cal M}^{(0)}(l_i^{\alpha_1\alpha_2};p_{i+1},\ldots,p_{j};-l_j^{\alpha_1\alpha_2})\times
{\cal M}^{(0)}(l_j^{\alpha_1\alpha_2};p_{j+1},\ldots,p_{k};-l_k^{\alpha_1\alpha_2})\nn &\times&
{\cal M}^{(0)}(l_k^{\alpha_1\alpha_2};p_{k+1},\ldots,p_{i};-l_i^{\alpha_1\alpha_2})\ ,
\eeqa
with an implicit sum over the internal states of the cut lines.
For example, the residue of the amplitude  for the pure 6-gluon amplitude with $d_2=d_3=d_4=0$ factorizes into (see fig.~\ref{fig:fig3})
\beqa
\mbox{Res}_{234}\Big({\cal A}_6(l^{\alpha_1\alpha_2})\Big)&=&
{\cal M}_3^{(0)}(l_2^{\alpha_1\alpha_2};p_3;-l_3^{\alpha_1\alpha_2})\times
{\cal M}_3^{(0)}(l_3^{\alpha_1\alpha_2};p_4;-l_4^{\alpha_1\alpha_2})\nn &\times&
{\cal M}_6^{(0)}(l_4^{\alpha_1\alpha_2};p_5,p_6,p_1,p_2;-l_2^{\alpha_1\alpha_2})\ .
\eeqa

The remaining dependence of the residue $\overline{c}_{ijk}$ on the loop momentum 
resides in the trivial space
\beq
\overline{c}_{ijk}(l)\equiv\overline{c}_{ijk}(s_1,s_2);\ s_1=n_1\cdot l,\, s_2=n_2\cdot l\ .
\eeq
The maximum rank of the triangle diagrams in standard model processes is 3. This gives us 10 possible terms 
($\{1,s_1,s_2,s_1^2,s_1s_2,s_2^2,s_1^3,s_1^2s_2,s_1s_2^2,s_2^3\}$) for the most general polynomial
form of the residue. However using Eq.~(\ref{metricdecomp}) we have the constraint $s_1^2+s_2^2\sim n_1^2+n_2^2=2$ which 
reduces the number of terms to 7. The form we chose is given by
\beq\label{eqn:Cspurious}
\overline{c}_{ijk}(l)=c_{ijk}^{(0)}+c_{ijk}^{(1)}s_1+c_{ijk}^{(2)}s_2+c_{ijk}^{(3)}(s_1^2-s_2^2)
+s_1s_2(c_{ijk}^{(4)}+c_{ijk}^{(5)}s_1+c_{ijk}^{(6)}s_2)\ .
\eeq
We need to determine all 7 constants $c_{ijk}^{(n)}$ by constructing 7 equations. This is accomplished
by choosing 7 combinations of $(\alpha_1,\alpha_2)$ with $\alpha_1^2+\alpha_2^2=-(V_3^2-m_k^2)$ in Eq.~(\ref{trianglecoeff}). 
This system of equations 
can be easily solved using a matrix inversion. Note that in principle
we can generate an unlimited set of equations. This can be useful in a numerical application to obtain
better numerical accuracy.

With the above prescription it is now easy to determine the  spurious term
for any value of the loop momentum. Finally we note that the integration over the term
\beq\label{Cdef3}
\int [d\,l]\ \frac{\overline{c}_{ijk}(l)}{d_id_jd_k}=
c^{(0)}_{ijk}\int [d\,l]\ \frac{1}{d_id_jd_k}=c_{ijk}I_{ijk}\ ,
\eeq
is now trivially done, giving us the triangle coefficient times the triangle master integral.

\subsection{Construction of the bubble residue}

This sub-section follows closely the previous two sub-sections.
To calculate the bubble coefficients we need to put two propagators on-shell. The box
and triangle contributions need to be explicitly subtracted. This gives for the bubble coefficient
\beq\label{bubblecoeff}
\overline{b}_{ij}(l)=\mbox{Res}_{ij}
\left({\cal A}_N(l)-\sum_{k\neq i,j}\frac{\overline{c}_{ijk}(l)}{d_id_jd_k}
-\frac{1}{2!}\sum_{k,l\neq i,j}\frac{\overline{d}_{ijkl}(l)}{d_id_jd_kd_l}\right)\ .
\eeq
Decomposing the loop momentum in the NV-basis of the two inflow momenta with $d_i=d_j=0$ 
(choosing $q_j=0$) gives us
according to Eq.~(\ref{eq:l3})
\beq\label{Ldef2}
l^\mu=V_2^\mu+\alpha_1 n_1^\mu+\alpha_2 n_2^\mu+\alpha_3 n_3^\mu\ ,
\eeq
with
\beq\label{Vdef2}
V_2^{\mu}=
-\frac{1}{2}(q_{i}^2-m_{i}^2+m_{j}^2)\, v_1^{\mu}\ ,
\eeq
where
\beq\label{videf2}
v_1^{\mu}=\frac{\delta^{\mu}_{k_1}}{\Delta(k_1)}=\frac{k_1^\mu}{k_1^2}\ ,
\eeq
and
\beq
k_1=q_i\ .
\eeq
The base vectors of the trivial space $\{n_1,n_2,n_3\}$ have to be explicitly constructed
using the constraints
\beq\label{Ndef2}
n_i\cdot n_j=\delta_{ij};\ n_i\cdot k_j=0;\ w^{\mu\nu}(k_1,k_2)=n_1^\mu n_1^\nu+n_2^\mu n_2^\nu+n_3^\mu n_3^\nu\ .
\eeq
The solution to the unitarity constraints ($d_i=d_j=0$) gives as a infinite set of solutions
\beq
l_{\alpha_1\alpha_2\alpha_3}^{\mu}=V_2^\mu+\alpha_1\,n_1^\mu+\alpha_2\,n_2^\mu+\alpha_3\,n_3^\mu;\ 
\alpha_1^2+\alpha_2^2+\alpha_3^2=-(V_2^2-m_j^2)\ .
\eeq
\begin{figure}[t!]
   \begin{center}
   \leavevmode
   \epsfysize=8.0cm
   \centerline{\epsfig{figure=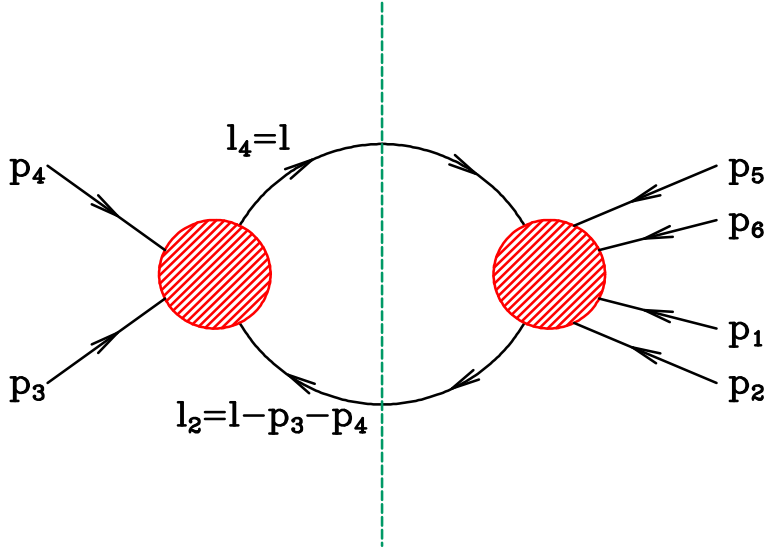,angle=-90,width=10cm}}
\end{center}
\caption{The factorization of the 6-gluon amplitude for the calculation of the $\overline{b}_{24}(l)$ residue
with the loop momentum parametrization choice $q_0=-p_5-p_6=p_1+p_2+p_3+p_4$. }
\label{fig:fig4}
\end{figure}
As before the residue of the amplitude  factorizes into tree-level amplitudes, in this case 
into 2 tree-level amplitudes with an implicit sum over the states of the cut lines 
\beqa\lefteqn{
\mbox{Res}_{ij}\Big({\cal A}_N(l^{\alpha_1\alpha_2\alpha_3})\Big)=} \nn &&
{\cal M}^{(0)}(l_i^{\alpha_1\alpha_2\alpha_3};p_{i+1},\ldots,p_{j};-l_j^{\alpha_1\alpha_2\alpha_3})\times
{\cal M}^{(0)}(l_j^{\alpha_1\alpha_2\alpha_3};p_{j+1},\ldots,p_{k};-l_k^{\alpha_1\alpha_2\alpha_3})\ .
\eeqa
For example, the residue of the amplitude  for the pure 6-gluon amplitude with $d_2=d_4=0$ factorizes into (see fig.~\ref{fig:fig4})
\beq
\mbox{Res}_{24}\Big({\cal A}_6(l^{\alpha_1\alpha_2\alpha_3})\Big)=
{\cal M}_4^{(0)}(l_2^{\alpha_1\alpha_2\alpha_3};p_3,p_4;-l_4^{\alpha_1\alpha_2\alpha_3})\times
{\cal M}_6^{(0)}(l_4^{\alpha_1\alpha_2\alpha_3};p_5,p_6,p_1,p_2;-l_2^{\alpha_1\alpha_2\alpha_3})\ .
\eeq

The remaining loop dependence of the residue $\overline{b}_{ij}$ is in the trivial space
\beq
\overline{b}_{ij}(l)\equiv\overline{b}_{ij}(s_1,s_2,s_3);\ s_1=n_1\cdot l,\, s_2=n_2\cdot l,\, s_3=n_3\cdot l\ .
\eeq
The maximum rank of the bubble in standard model processes is 2. This gives us 10 possible terms 
($\{1,s_1,s_2,s_3,s_1^2,s_2^2,s_3^2,s_1s_2,s_1s_3,s_2s_3\}$) for the most general polynomial
form of the residue. However, using Eq.~(\ref{metricdecomp}) we have the constraint 
$s_1^2+s_2^2+s_3^2\sim n_1^2+n_2^2+n_3^2=3$ which 
reduces the number of terms to 9. The form we chose is given by
\beq
\overline{b}_{ij}(l)=b_{ij}^{(0)}+b_{ij}^{(1)}s_1+b_{ij}^{(2)}s_2+b_{ij}^{(3)}s_3
+b_{ij}^{(4)}(s_1^2-s_3^2)+b_{ij}^{(5)}(s_2^2-s_3^2)+b_{ij}^{(6)}s_1s_2+b_{ij}^{(7)}s_1s_3
+b_{ij}^{(8)}s_2s_3\ .
\eeq
We need to determine all 9 constants $b_{ij}^{(n)}$ by constructing 9 equations. This is accomplished
by choosing 9 combinations of $(\alpha_1,\alpha_2,\alpha_3)$ with
 $\alpha_1^2+\alpha_2^2+\alpha_3^2=-(V_2^2-m_j^2)$ 
in Eq.~(\ref{bubblecoeff}).  This system of equations
can be easily solved using a matrix inversion. Note that in principle
we can generate an unlimited set of equations. This can be useful in a numerical application to obtain
better numerical accuracy. We only need the full bubble residue in the case when 
the tadpole contribution is non-zero.

With the above prescription it is now easy to determine the spurious term 
for any value of the loop momentum. Finally we note that the integration over the term
\beq\label{Cdef2}
\int d\,l\ \frac{\overline{b}_{ij}(l)}{d_id_j}=
b^{(0)}_{ij}\int d\,l\ \frac{1}{d_id_j}=b_{ij}I_{ij}\ ,
\eeq
is now trivially done, giving us the bubble coefficient times the bubble master integral.

\subsection{Construction of the tadpole coefficient}

In the case there is a tadpole contribution we need to determine its coefficient from the
relation 
\beq
\overline{a}_{i}(l)=\mbox{Res}_i
\left({\cal A}_N(l)-\sum_{j\neq i}\frac{\overline{b}_{ij}(l)}{d_id_j}
-\frac{1}{2!}\sum_{j,k\neq i}\frac{\overline{c}_{ijk}(l)}{d_id_jd_k}
-\frac{1}{3!}\sum_{j,k,l\neq i}\frac{\overline{d}_{ijkl}(l)}{d_id_jd_kd_l}\right)\ .
\eeq
Using Eq.~(\ref{eq:l3}) the loop momentum can be decomposed in four orthonormal vectors
\beq
l^\mu=\alpha_1 n_1^\mu+\alpha_2 n_2^\mu+\alpha_3 n_3^\mu+\alpha_4 n_4^\mu\ ,
\eeq
and
\beq
n_i\cdot n_j=\delta_{ij};\ g^{\mu\nu}=n_1^\mu n_1^\nu+n_2^\mu n_2^\nu+n_3^\mu n_3^\nu+n_4^\mu n_4^\nu\ .
\eeq
The solution to the unitarity constraints ($d_i=0$ with $q_i=0$) gives an infinite set of solutions
\beq
\alpha_1^2+\alpha_2^2+\alpha_3^2+\alpha_4^2=m_i^2 \ ,
\eeq
\begin{figure}[t!]
   \begin{center}
   \leavevmode
   \epsfysize=8.0cm
   \centerline{\epsfig{figure=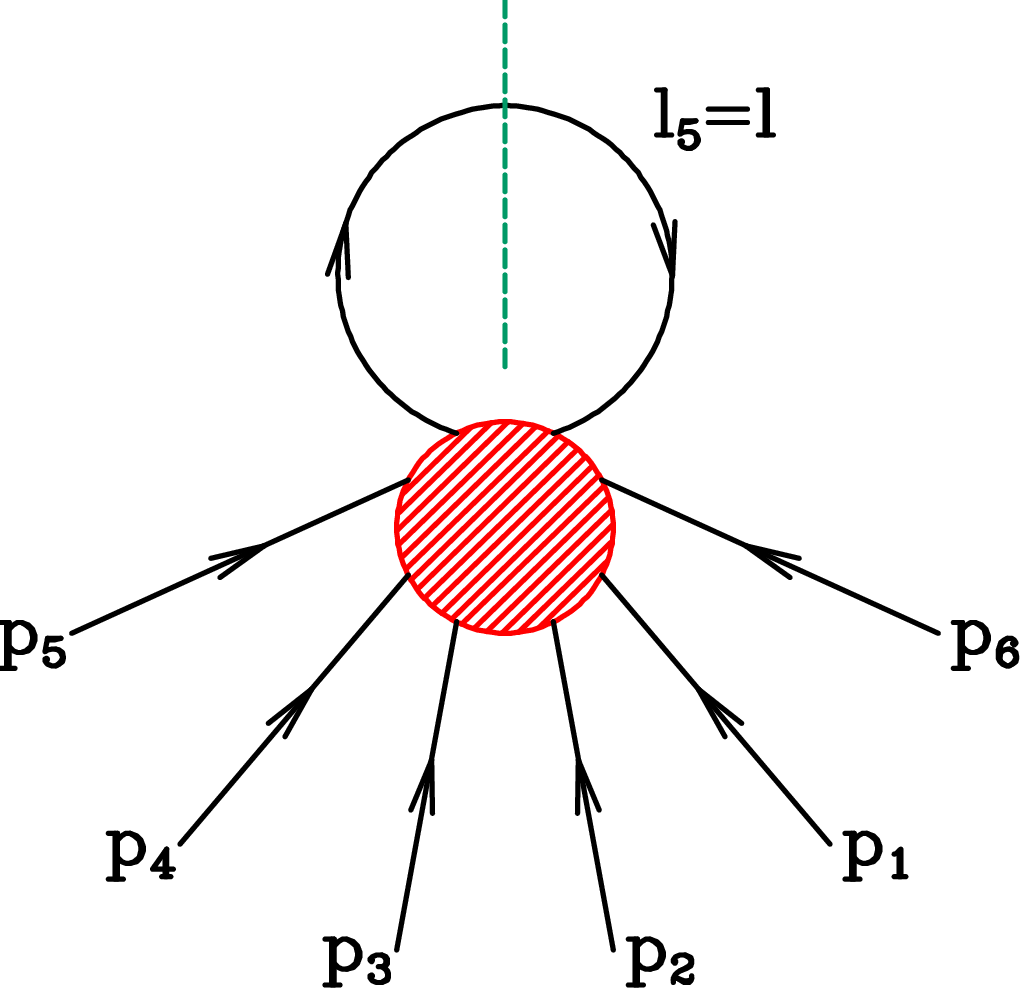,angle=-90,width=10cm}}
\end{center}
\caption{The factorization of the 6-gluon amplitude for the calculation of the $\overline{a}_{5}(l)$ residue
with the loop momentum parametrization choice $q_0=-p_6=p_1+p_2+p_3+p_4+p_5$. }
\label{fig:fig5}
\end{figure}
The residue of the amplitude  becomes a tree-level amplitude with an implicit sum over the states of the cut particle
\beq
\mbox{Res}_i\Big({\cal A}_N(l^{\alpha_1\alpha_2\alpha_3\alpha_4})\Big)=
{\cal M}^{(0)}(l_i^{\alpha_1\alpha_2\alpha_3\alpha_4};p_{i+1},\ldots,p_{i};-l_i^{\alpha_1\alpha_2\alpha_3\alpha_4})\ .
\eeq
For example, the residue of the amplitude  for the pure 6-gluon amplitude with $d_5$ factorizes into (see fig.~\ref{fig:fig5})
\beq
\mbox{Res}_5\Big({\cal A}_6(l^{\alpha_1\alpha_2\alpha_3\alpha_4})\Big)=
{\cal M}_8^{(0)}(l_5^{\alpha_1\alpha_2\alpha_3\alpha_4};p_6,p_1,p_2,p_3,p_4,p_5;-l_5^{\alpha_1\alpha_2\alpha_3\alpha_4})\ .
\eeq

The maximum rank of the tadpole in the Standard Model is one,
giving for the spurious term
\beq
\overline{a}_i =
a_i^{(0)} + a_i^{(1)} s_1 + a_i^{(2)} s_2 + a_i^{(3)} s_3 + a_i^{(4)} s_4\ .
\eeq

The coefficient $a_{i}$ of the master integral $I_i$ is easily obtained by e.g.
\beq
a_i^{(0)}=\frac{\overline{a}_i(l^{\alpha_1\alpha_2\alpha_3\alpha_4})
               +\overline{a}_i(l^{\beta_1\alpha_2\alpha_3\alpha_4})}{2}\ ,
\eeq
with $\alpha_1=-\beta_1=m_i$ and $\alpha_2=\alpha_3=\alpha_4=0$.

\section{Numerical results}

As an application we calculate the 4, 5 and 6 gluon scattering
amplitudes at one-loop using the method of sec. 2. 
The cut-constructible parts of the ordered amplitudes are also known analytically 
(\cite{Ellis:1985er,Bern:1990cu,Kunszt:1993sd}, \cite{Bern:1993mq},
\cite{Bern:1994zx,BDKOneloopInt,Bidder:2004tx,Bern:2005ji,Bern:2005cq,Britto:2005ha,Britto:2006sj}),
making a direct comparison possible. These multi-gluon scattering amplitudes
form a good test for numerical procedures as they are the sum over a
large number of Feynman graphs with significant gauge
cancellations. Also, the 6-gluon amplitude was numerically evaluated
using the integration-by-parts method \cite{Ellis:2006ss}\footnote{This Feynman diagram
calculation yielded the full amplitude including rational terms.}. The
numerical evaluation time using that method was around 9 seconds per
ordered amplitude (on a 2.8GHz Pentium processor). We can compare this
evaluation time with the unitarity method of the previous section.
This directly compares the computational effort between a numerical
method using Feynman diagrams with form factor expansion
and the numerical unitarity method using analytical expressions
for the tree-level amplitudes.

To calculate the gluon scattering amplitude we need to determine all the master integral coefficients and
combine these with the master integrals according to Eq.~(\ref{MasterDecomp}). It is 
straightforward to numerically evaluate a coefficient using the method outlined in the previous section. 
Given a set of external momenta we simply calculate the 
appropriate four vectors $\{v_i\}$ and $\{n_i\}$ 
according to Eqs.~(\ref{videf4}, \ref{videf3}, \ref{Ndef3}, \ref{videf2}, \ref{Ndef2}). 
From these vectors we construct the special loop momenta of Eqs.~(\ref{Ldef4}, \ref{Ldef3}, \ref{Ldef2})
which sets the appropriate denominators to zero. Using the analytically known leading order gluon amplitudes
we can calculate the coefficient of Eqs.~(\ref{Cdef4}, \ref{Cdef3}, \ref{Cdef2}). Note that the multi-gluon scattering
amplitudes do not get a contribution from the tadpole master integrals.
We therefore need only to calculate the $b_{ij}^{(0)}$-coefficient of the bubble residue and not the remaining
8 coefficients of the spurious term. 

The first check on the numerical implementation is performed by calculating the $\epsilon^{-2}$ term,
where $D=4-2 \epsilon$. 
This term gets contributions
both from the box master integrals and from triangle master integrals with one leg off-shell. 
The $\epsilon^{-2}$-term for a $n$-gluon
ordered amplitude is proportional to the leading-order amplitude and is given by
\beq
m^{(1)}(1,2,\ldots,n)\sim -\frac{n}{\epsilon^2}\times m^{(0)}(1,2,\ldots,n)+{\cal O}\left(\epsilon^{-1}\right)\ .
\eeq

The next check is the $\epsilon^{-1}$ term. Again, the term is proportional to the leading-order ordered amplitude. This term 
gets contributions from the one-, two- and three-leg off-shell box master integrals, 
one- and two-leg off-shell triangle master integrals and bubble master integrals. 
The contribution is given by
\beq
m^{(1)}(1,2,\ldots,n)\sim \left(-\frac{n}{\epsilon^2}
+\frac{1}{\epsilon}\left(-\frac{11}{3}+\sum_{i=1}^n \log\left(\frac{s_{i,i+1}}{\mu^2}\right)\right)\right)\times m^{(0)}(1,2,\ldots,n)
+{\cal O}\left(1\right)\ ,
\eeq
where $s_{i,i+1}=2 p_i \cdot p_{i+1}$ and $s_{n,n+1}=s_{n,1}$. As usual $\mu$ is the scale 
introduced to maintain the dimension of the integral in $4-2\epsilon $ dimensions.

When the implementation passes both non-trivial tests we have checked all coefficients proportional to infrared and collinear
divergent master integrals. This leaves only the 3-leg off-shell triangle coefficients and the 4-leg off-shell box coefficients unchecked.
Note that for the 4-gluon and 5-gluon ordered amplitudes there are no finite master integral contributions and hence all
coefficients are checked by looking at the divergent parts. The 6-gluon amplitude gets a contribution from the finite 3-leg
off-shell triangle master integral.

To compare with the analytic results we generate 100,000 flat phase space events 
for the $2\rightarrow (n-2)$ gluon scattering using RAMBO~\cite{Kleiss:1985gy}. 
The center-of-mass collision energy is $\sqrt{S}$.
The events are required to have the following cuts on the outgoing gluons:  
a cut on the transverse energy, $E_T>0.01\times \sqrt{S}$,
a maximum rapidity, $\eta<3$ and a separation cut, $\Delta R > 0.4$.

\begin{figure}[t!]
   \begin{minipage}{0.3\linewidth}
      \centerline{\epsfig{figure=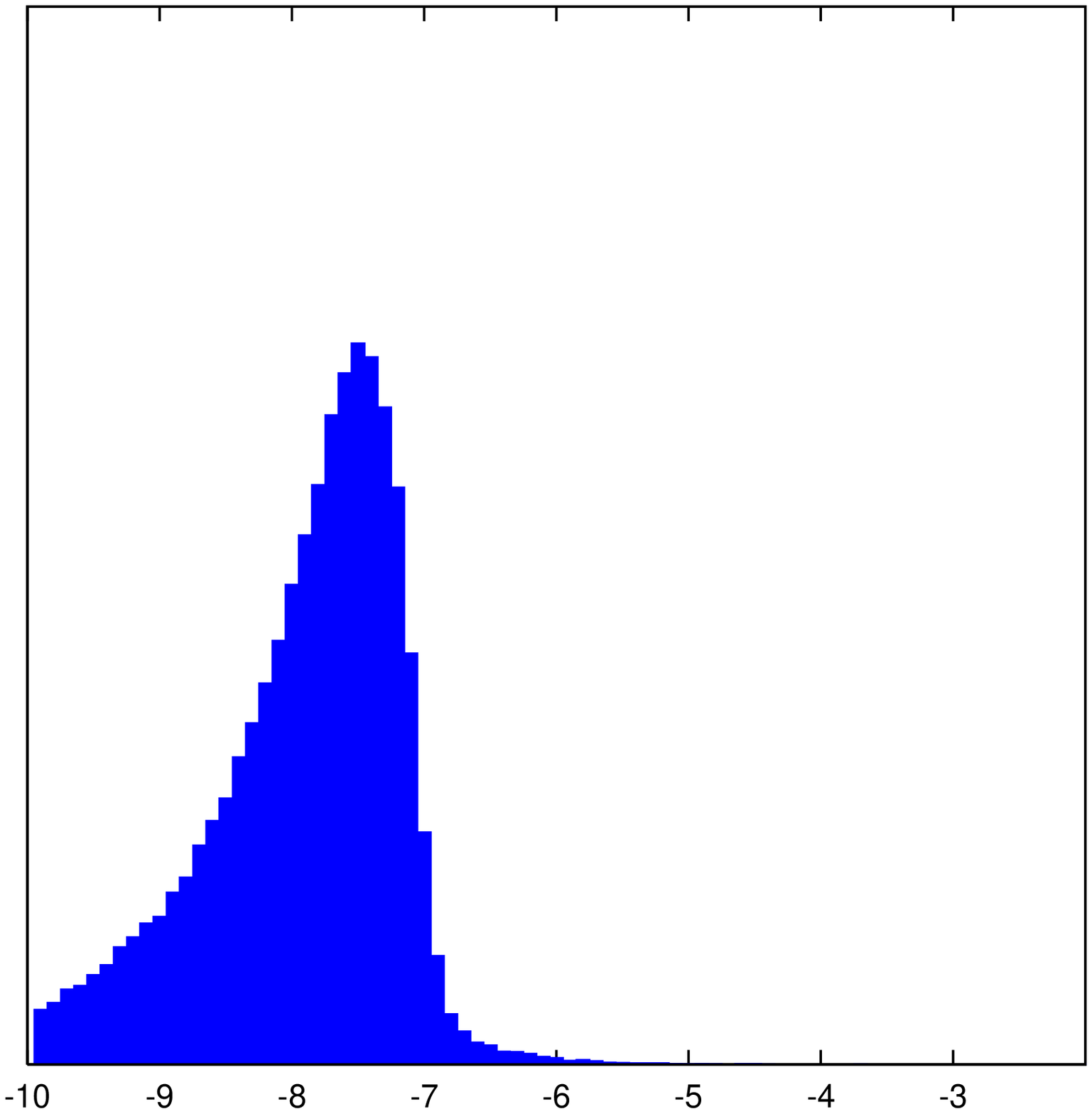,angle=-90,width=1.61\linewidth}}
   \end{minipage}		
   \begin{minipage}{0.3\linewidth}
      \centerline{\epsfig{figure=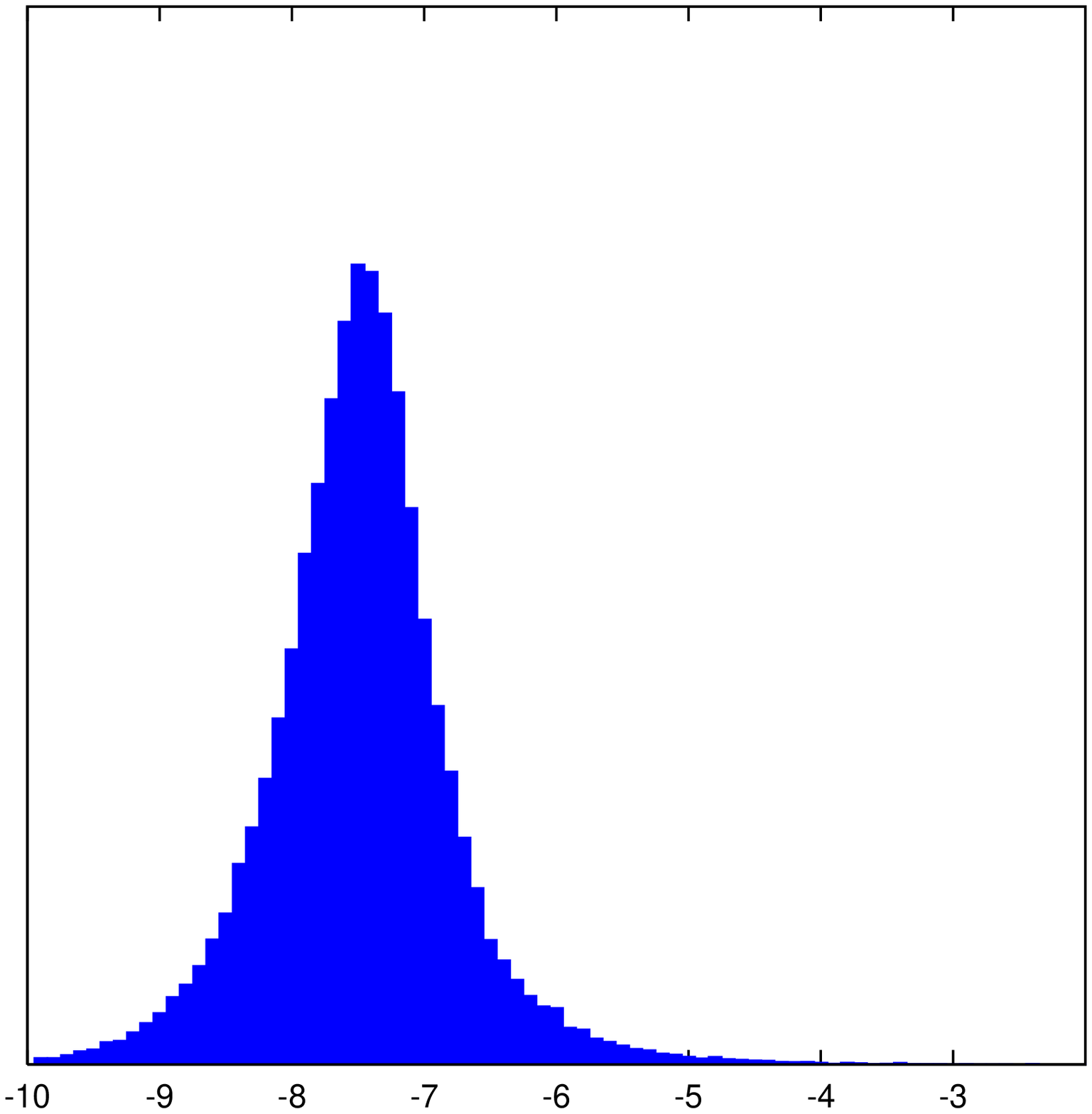,angle=-90,width=1.61\linewidth}}
   \end{minipage}		
   \begin{minipage}{0.3\linewidth}
      \centerline{\epsfig{figure=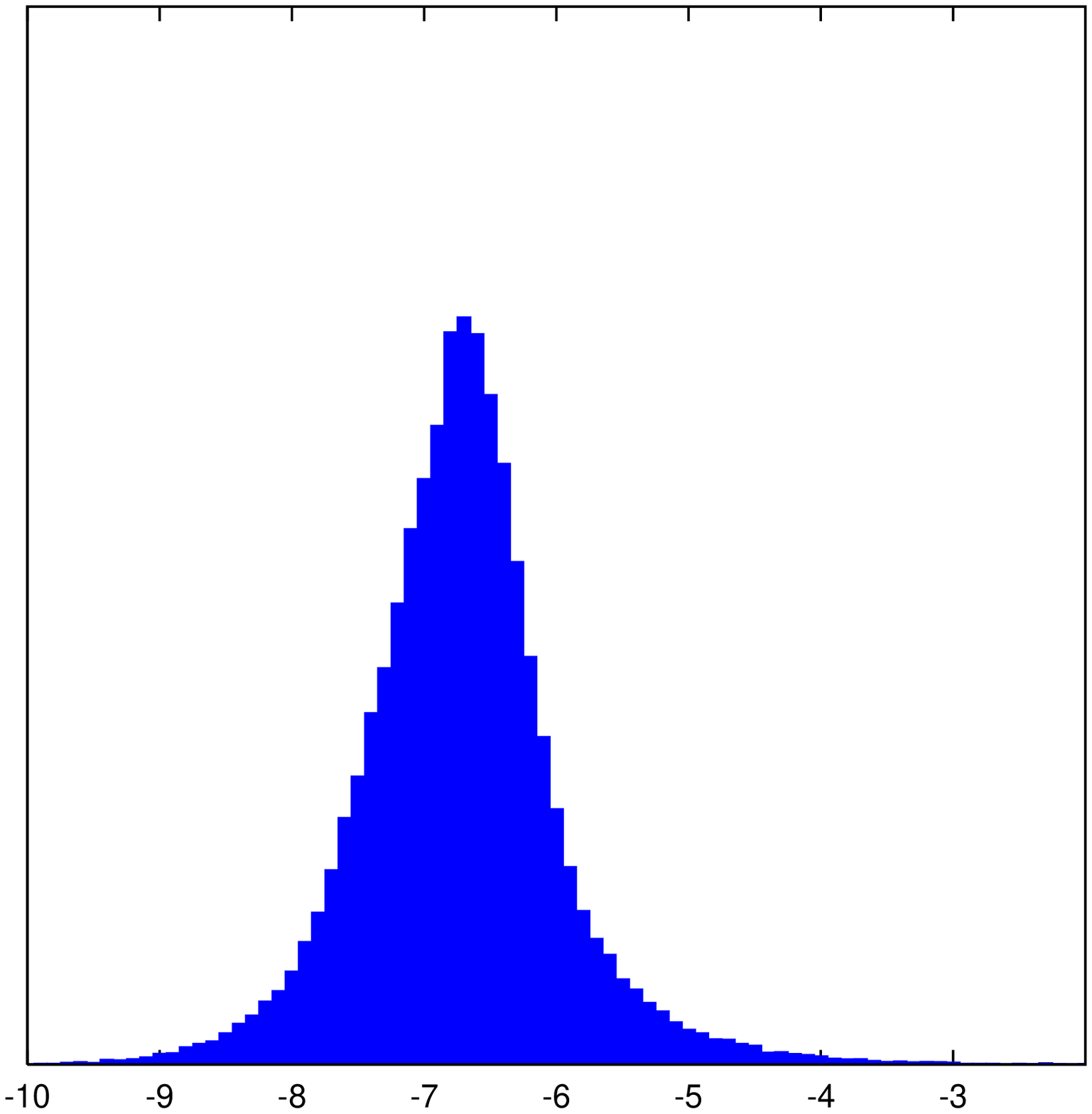,angle=-90,width=1.61\linewidth}}
   \end{minipage}		
\caption{The relative error for 100,000 ordered 4-gluon amplitude for the $(++--)$ helicity choice. 
The horizontal axis is the log-10 of the relative error of 
Eq.~(\ref{significance}), the vertical axis is the number of 
events in arbitrary linear units.
The left figure is the $\epsilon^{-2}$ contribution,
the middle figure is the $\epsilon^{-1}$ contribution and the right figure is the finite part.}
\label{fig:4-gluon}
\end{figure}
The evaluation time for 10,000 events is: for a $2\rightarrow 2$ gluon ordered helicity amplitude 9 seconds,
for a $2\rightarrow 3$ gluon ordered helicity amplitude 35 seconds and for a
for a $2\rightarrow 4$ gluon ordered helicity amplitude 107 seconds.
Note that using the integration-by-parts method of ref.~\cite{Ellis:2006ss} 
the evaluation time for 10,000 events would be approximately 90,000 second.
This means the unitarity method of section 2 improves the evaluation of the six-gluon amplitudes by 
a factor of approximately 900, almost 3 orders of magnitude.
The six-gluon evaluation is only three times slower than the five gluon
evaluation and eleven times slower than the four gluon amplitude.  This can be understood by counting the number
of coefficients needed to evaluate the scattering amplitude. The number of coefficients multiplying
a non-zero master integral for the $n$-gluon scattering ordered amplitude is
\beq
\left(\begin{array}{c} n \\ 4 \end{array}\right)+\left(\begin{array}{c} n \\ 3 \end{array}\right)
+\left(\begin{array}{c} n \\ 2 \end{array}\right) - n=\frac{n}{24}\left(n^3-2n^2+11n-34\right)\ ,
\eeq
\begin{figure}[t!]
   \begin{minipage}{0.3\linewidth}
      \centerline{\epsfig{figure=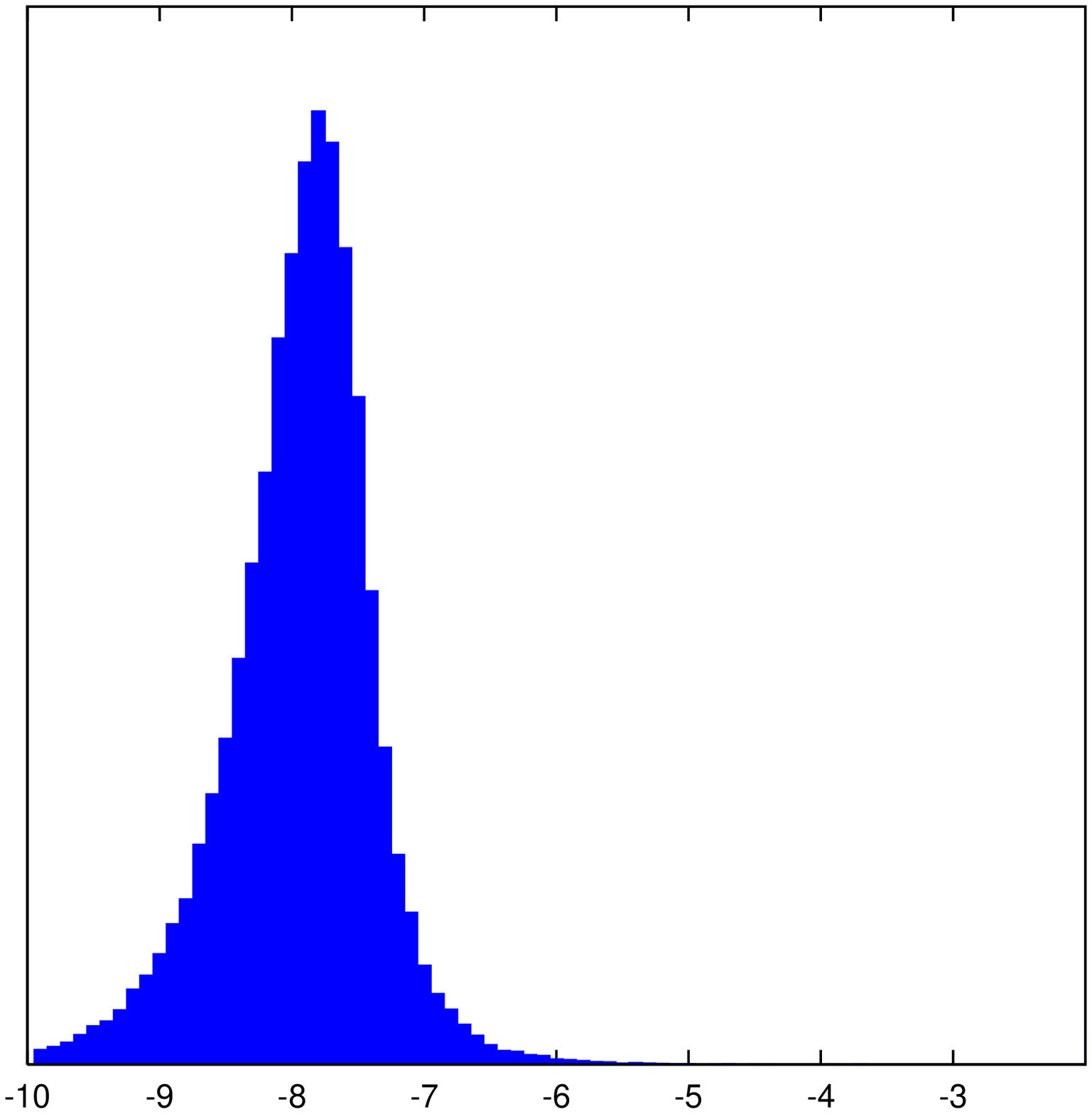,angle=-90,width=1.61\linewidth}}
   \end{minipage}		
   \begin{minipage}{0.3\linewidth}
      \centerline{\epsfig{figure=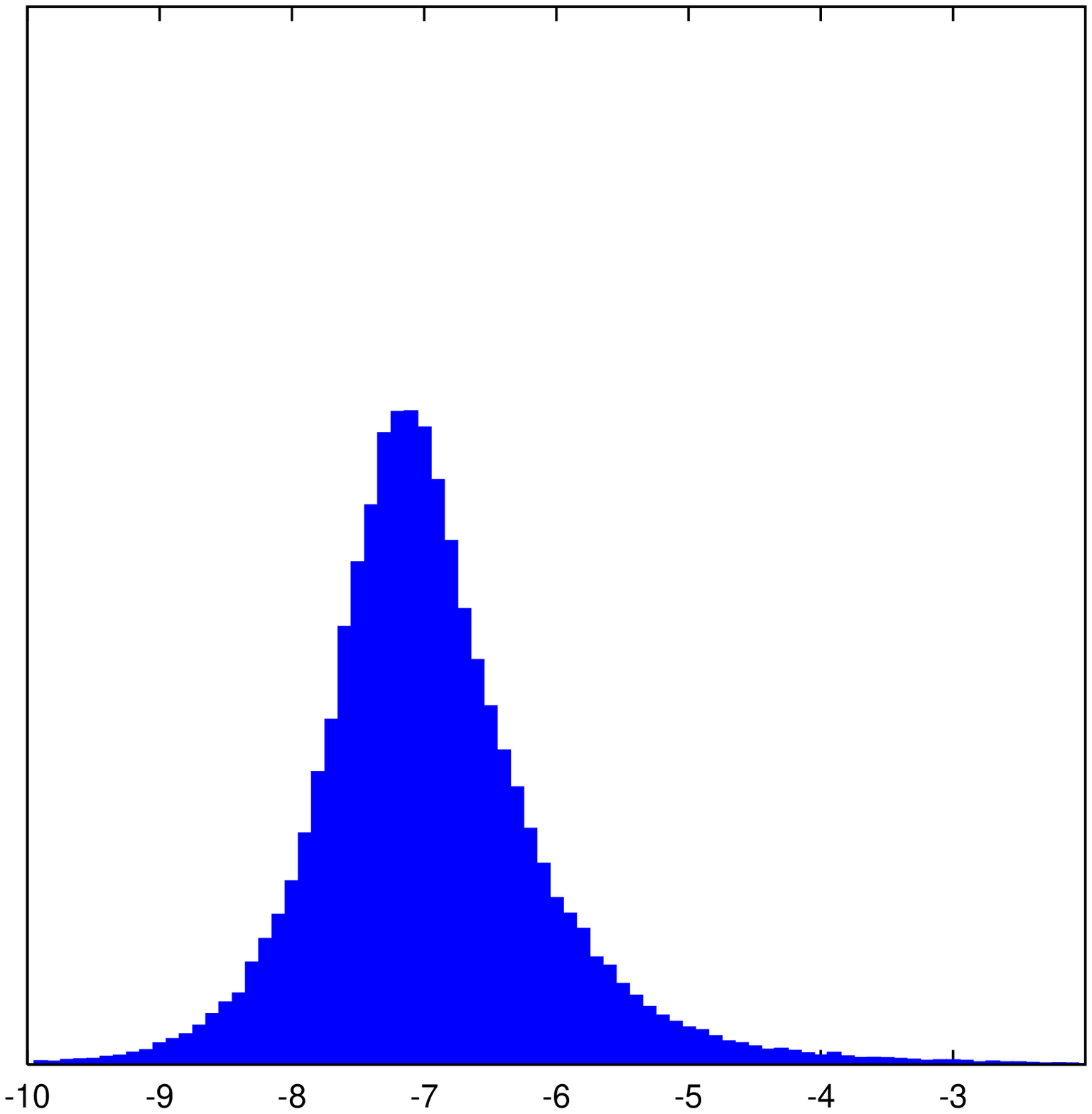,angle=-90,width=1.61\linewidth}}
   \end{minipage}		
   \begin{minipage}{0.3\linewidth}
      \centerline{\epsfig{figure=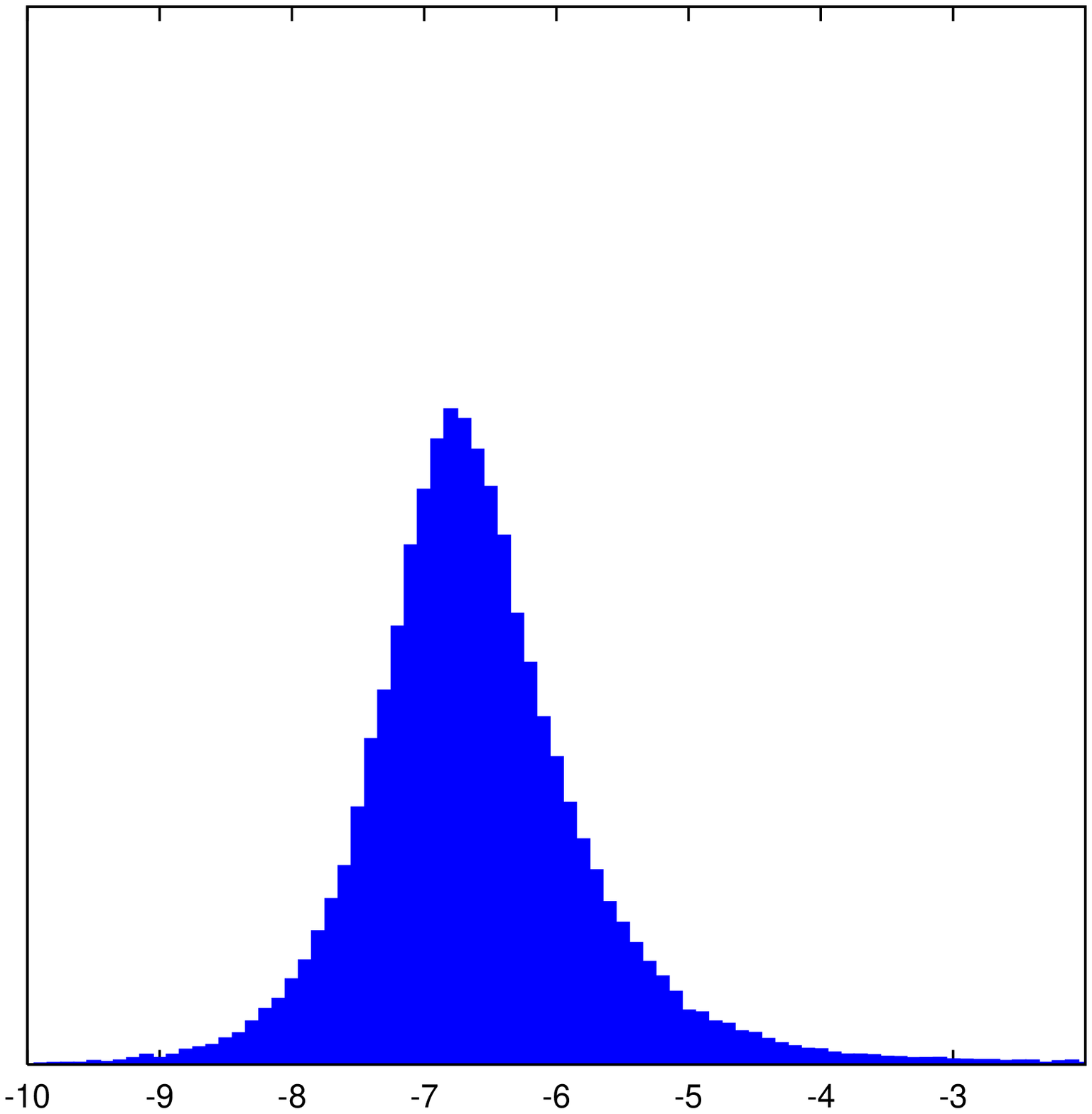,angle=-90,width=1.61\linewidth}}
   \end{minipage}		
\caption{Same as fig. \ref{fig:4-gluon}, but the ordered 5-gluon amplitude for the $(++---)$ helicity choice.}
\label{fig:5-gluon}
\end{figure}
where the first term is the number of box coefficients, the second term the number of triangle coefficients 
and the third term gives the number of self energy coefficients. Finally, the last term subtracts the $n$ external
gluon bubble master integrals (because the corresponding master integral is zero). That is, the number of 4-gluon coefficients is six,
the number of 5-gluon coefficients is twenty and the number of 6-gluon coefficients forty-four. 
The computational time
roughly follows the number of coefficients to be calculated for the scattering. This is a very different scaling law
than the $n!$ growth of a straightforward Feynman diagram expansion
(as is the case for the integration-by-part algorithm of ref. \cite{Ellis:2006ss}).
The ``factorization'' of the diagrams into the tree-level blobs explains this large difference and causes
the large gain in speed using the unitarity method compared to more conventional methods, replacing a
$ n!$ factorial growth by a $n^4$ power growth. The advantage of unitarity method with respect to a 
a Feynman-diagram based approach becomes greater as the number of 
external particles grows (provided the tree-level amplitudes 
are known).

\begin{figure}[t!]
   \begin{minipage}{0.3\linewidth}
      \centerline{\epsfig{figure=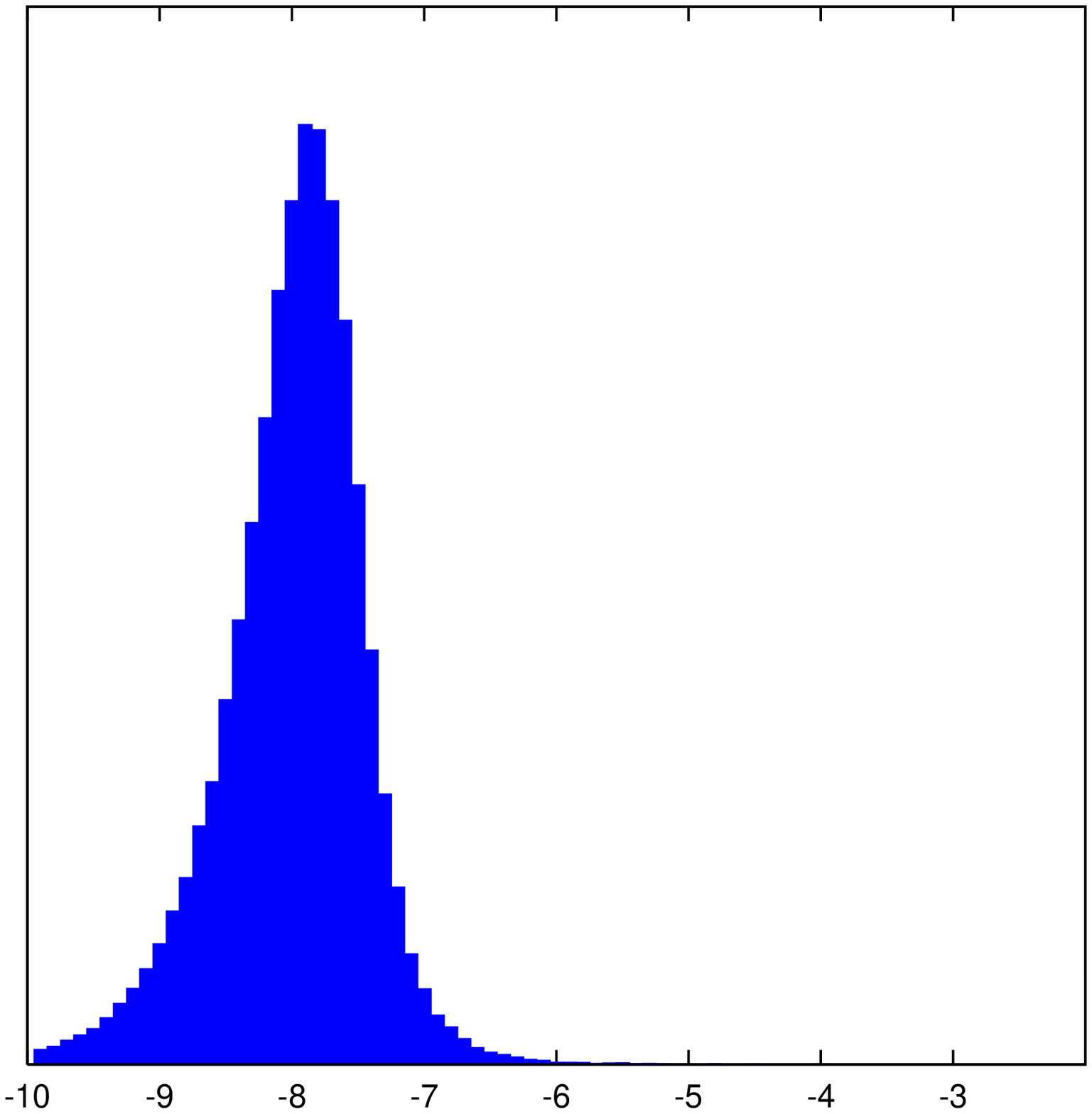,angle=-90,width=1.61\linewidth}}
   \end{minipage}		
   \begin{minipage}{0.3\linewidth}
      \centerline{\epsfig{figure=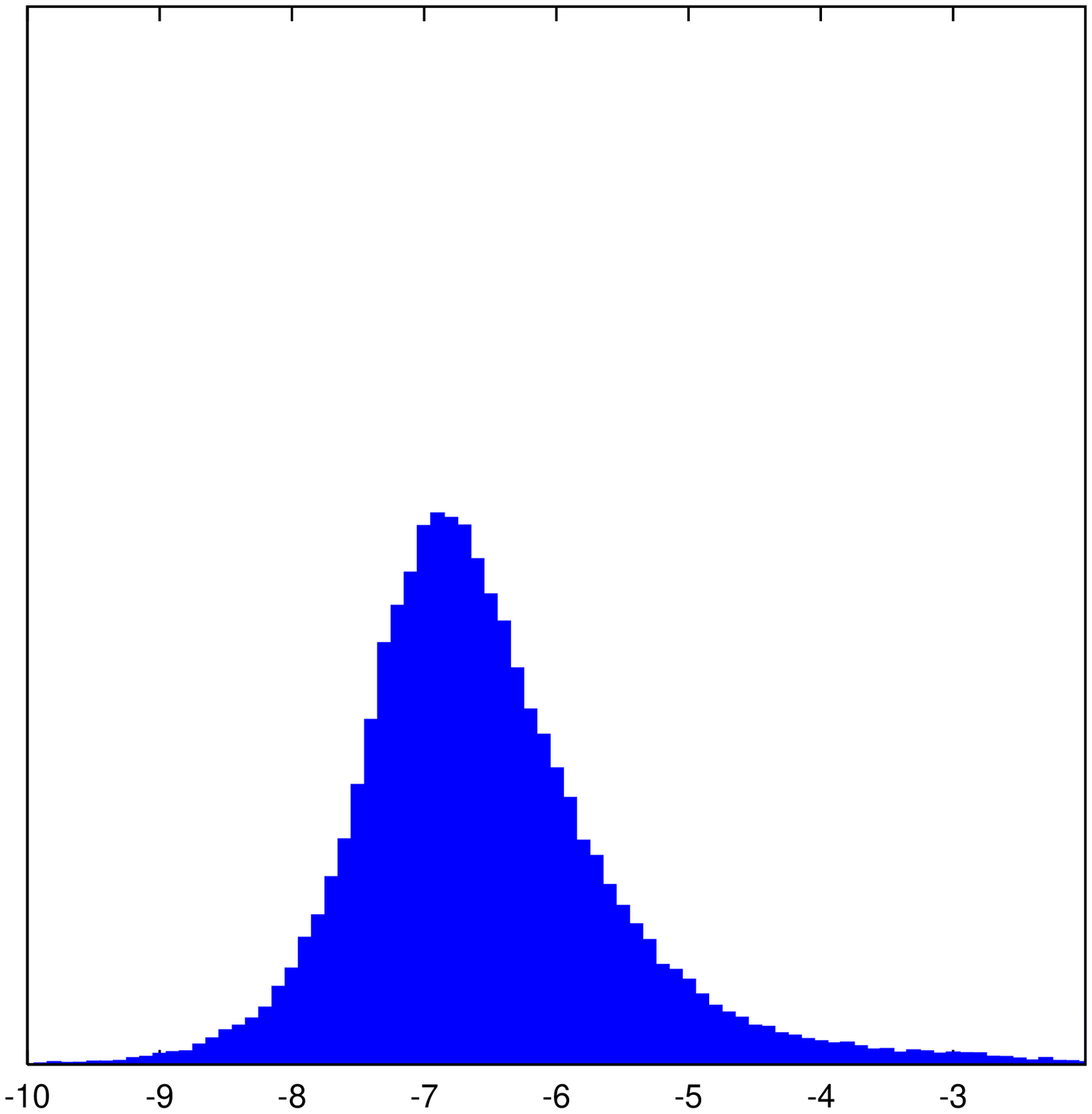,angle=-90,width=1.61\linewidth}}
   \end{minipage}		
   \begin{minipage}{0.3\linewidth}
      \centerline{\epsfig{figure=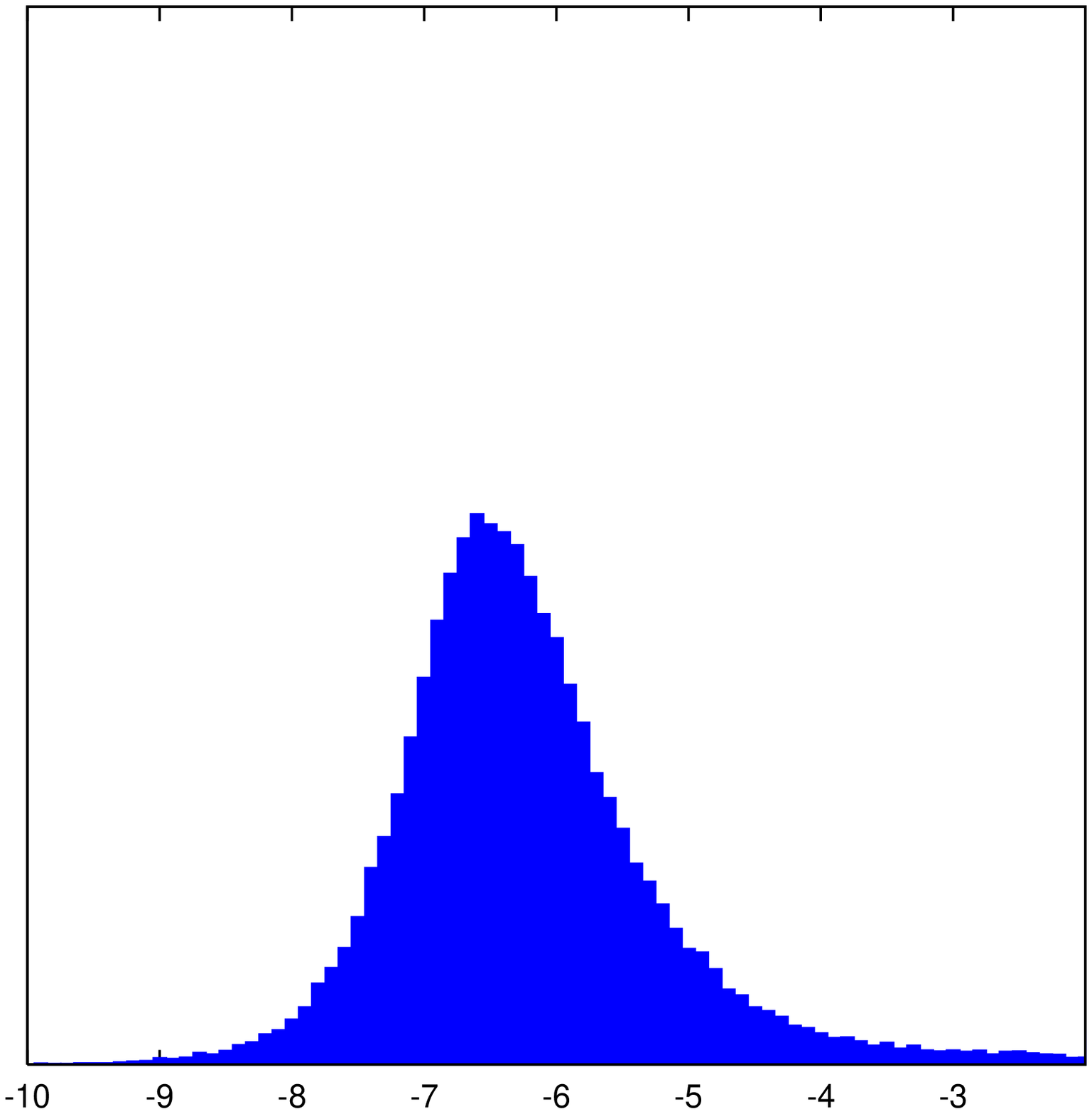,angle=-90,width=1.61\linewidth}}
   \end{minipage}		
\caption{Same as fig. \ref{fig:4-gluon}, but the ordered 6-gluon amplitude for the $(++----)$ helicity choice.}
\label{fig:6-gluon}
\end{figure}

The numerical comparisons for the cut constructible part of $m_4(++--)$, $m_5(++---)$ and $m_6(++----)$ ordered
helicity amplitudes
are summarized in figs. (\ref{fig:4-gluon},\ref{fig:5-gluon},\ref{fig:6-gluon})
\footnote{We also compared all other helicity combinations with the known results in the literature, leading to similar results.}. 
The degree of agreement is quantified by the expression
\beq\label{significance}
S=\log_{10}\left(\left|\frac{m^{(1)}_{\mbox{unitarity}}-m^{(1)}_{\mbox{analytic}}}{m^{(1)}_{\mbox{analytic}}}\right|\right)\ .
\eeq
In the 3 figures the 100,000 events are compared and binned in the quantity $S$ for each of the 3 contributions:
$\epsilon^{-2}$, $\epsilon^{-1}$ and finite. As can be seen the majority of the events agree with a relative precision
of $10^{-6}$ or better. This is more than sufficient for TEVATRON, LHC and ILC applications. However, for the $\epsilon^{-1}$
and finite parts a small portion of the events have a worse agreement. This is related to the 
fact that in these cases the amplitude receives a contribution from the
bubble coefficient. In particular it is related
to the calculation of the triangle subtraction term needed in the
calculation of the bubble coefficient. For a small fraction of the phase space points there is 
a numerical instability in the matrix inversion needed to
calculate the coefficients of the triangle spurious term in Eq.~(\ref{eqn:Cspurious}). 
To understand this better we simply rotate the two basis vectors of the trivial space of the triangle in the following manner
\beqa
n_+^{\mu}&=&n_1^{\mu}+i\,n_2^{\mu} \nn
n_-^{\mu}&=&n_1^{\mu}-i\,n_2^{\mu}\ ,
\eeqa
such that
\beq
n_+\cdot n_+=n_-\cdot n_-=0;\ n_+\cdot n_-=2\ .
\eeq
Within this basis the spurious term becomes
\beq
\overline{c}=c_0^{\prime}+c_1^{\prime}s_++c_2^{\prime}s_+^2+c_3^{\prime}s_+^3+c_4^{\prime}s_-+c_5^{\prime}s_-^2+c_6^{\prime}s_-^3\ ,
\eeq
where $s_+=l\cdot n_+$ and $s_-=l\cdot n_-$. By choosing 7 loop momenta $\{l_i\}_{i=1}^7$ we get the set of equations
\beq
m_i=\sum_{j=1}^7 A_{ij} c_j^{\prime}\Rightarrow c_j^{\prime}=\sum_{i=1}^7 m_i A_{ij}^{-1}\ ,
\eeq
where $m_i=\overline{c}(l_i)$ and $A_{ij}=\left(1,s_+(l_i),s_+^2(l_i),s_+^3(l_i),s_-(l_i),s_-^2(l_i),s_-^3(l_i)\right)$.
We note that the matrix $A_{ij}$ is a double vanderMonde-matrix \cite{NumRec}, i.e. a matrix of the form
\beq A_{ij}= \left(\begin{array}{ccccccc} 
1 &\alpha_1 &\alpha_1^2 &\alpha_1^3 &\beta_1&\beta_1^2&\beta_1^3 \\
1 &\alpha_2 &\alpha_2^2 &\alpha_2^3 &\beta_2&\beta_2^2&\beta_2^3 \\
\vdots &\vdots &\vdots &\vdots &\vdots &\vdots &\vdots \\
1 &\alpha_7 &\alpha_7^2 &\alpha_7^3 &\beta_7&\beta_7^2&\beta_7^3
\end{array}\right)\ ,
\eeq
which is known to give an unstable inverse. No numerical procedure is known to stabilize the calculation
of the inverse. In the figures we simply used the inverse anyway to see if these numerical issues would appear. We see
that for a small fraction of the events  they indeed appear 
in the case of the $\epsilon^{-1}$ and finite contributions. On the other hand
it is important to realize we have an infinite set of equations (i.e. an infinite set of loop momenta fulfilling the triple cut unitarity
condition) with only 7 coefficients to determine. This means we can achieve arbitrary precision using a $\chi^2$-type fitting procedure,
at the cost of more computer time. We did not pursue this method for this paper.

The final issue is the presence of Gram determinants in the box, triangle and bubble coefficients. We need to be aware of the 
Gram determinants when performing phase space integrals. We identify from the procedure in the last section two separate mechanisms
of generating Gram determinant type of denominator factors in the coefficient.

The first mechanism is straightforward. The solutions of the unitarity constraints for the box by quadruple cut gives a loop momentum
which is proportional to the inverse of the square root of the 3-particle Gram determinant as can be
seen in Eqs.~(\ref{Resdef4}-\ref{Vdef4})
\beq
\frac{\delta^{k_1k_2\mu}_{k_1k_2k_3}}{\Delta(k_1,k_2,k_3)}\sim\frac{1}{\sqrt{\Delta(k_1,k_2,k_3)}}\ .
\eeq
Because the maximum rank of the box tensor integral is 4, we can get at most terms of order $\Delta(k_1,k_2,k_3)^{-2}$. 
More precisely
\beq
d_{ijkl}\equiv A+\frac{B}{\Delta(k_1,k_2,k_3)^{1/2}}+\frac{C}{\Delta(k_1,k_2,k_3)}
+\frac{D}{\Delta(k_1,k_2,k_3)^{3/2}}+\frac{E}{\Delta(k_1,k_2,k_3)^2}\ ,
\eeq
with $k_1=(q_j-q_i)$, $k_2=(q_k-q_j)$ and $k_3=(q_l-q_k)$.
Similar, for the triangle and bubble coefficients we get respectively up to $\Delta(k_1,k_2)^{-3/2}$ and $\Delta(k_1)^{-1}$ terms.

The second source of Gram determinants is more subtle and is generated for 5-particle or higher scattering amplitudes.
It can happen that e.g.
$d_5(l_{1234})\rightarrow 0$ causing an instability at a specific phase space point. 
It should be treated with care when performing a numerical phase space integration.

\section{Summary and Outlook}

We have presented a formulation of the 4-dimensional unitarity cut method in a physical language using the
van Neerven-Vermaseren basis of ref.~\cite{van Neerven:1983vr}. This basis is also used
to reduce 5- (or higher) point tensor integrals to 4-point tensor integrals. When
applied to 4- (or lower) point tensor integrals 
the decomposition of the loop momentum generates extra basis vectors
spanning a ``trivial'' space. This trivial space is orthogonal to the external momenta of the loop integral. 
When applying unitarity cuts, the loop momentum dependence in the trivial space
generate so-called spurious terms~\cite{OPP}. These spurious terms integrate to zero when considering an individual cut
diagram. However, when combining the double, triple and quadruple cuts care has to be taken not to double count.
These spurious terms play an important role in the subtraction schemes needed to avoid the double counting problem.
The procedure outlined in this paper allows us to solve the
unitarity constraints without resorting to the explicit 4-dimensional spinor formalisms
used in analytic calculations. This makes the method equally applicable to processes with (complex) masses
for the internal lines.

When applying the 4-dimensional unitarity cuts to the amplitude, both the master integral coefficients and the spurious terms are
calculable in terms of factorized products of tree-level amplitudes. The cut lines cause two of the external 
momenta of the tree-level amplitude to be complex.
Existing leading-order generators, such as VECBOS~\cite{VECBOS} and NJETS~\cite{NJETS1,NJETS2}, can be upgraded to
allow for two complex external momenta without much effort. Once these upgraded tree-level generators are 
interfaced to the numerical program based on method described in this paper, they can be converted to one-loop generators without
any additional calculations.
This will allow us to automate the
calculation of  the cut-constructible part of the one-loop
amplitudes for many  important background processes at the TEVATRON and LHC  such as
$PP\rightarrow 4\  {\rm  or \  } 5\  {\rm  jets}$,\ $PP\rightarrow W\  + 3,\ {\rm or \  4}\   {\rm jets} $\,, \
$PP\rightarrow t + \bar{t} + 1\,, 2\  {\rm or }\  3\  {\rm jets} $\,, \
$PP\rightarrow t + \bar{t} + b + \bar{b} + 0\ {\rm or }\   1\
{\rm jets}$ \ (massless bottom quark) etc.\,.

As an example, we took the analytically known 4-, 5- and 6-gluon ordered tree-level helicity amplitudes 
and used them to evaluate the cut-constructible part of the 4-, 5- and 6-gluon ordered one-loop
helicity amplitudes. Especially the 6-gluon amplitudes are very complex, demonstrating the power
of the method in the paper. The computational time required to evaluate the amplitudes is fast
enough for serious TEVATRON, LHC and ILC applications. 
The numerical instabilities in certain phase space points due to 
additional linear dependences between the particle momenta (the ``Gram determinant instabilities'')
are easily identified in the master integral coefficients. The degree of the instabilities are
the same as one gets from analytic unitarity calculations.

The ultimate goal is to construct a NLO parton level generator. To reach that point, two outstanding challenges remain.
The first one is the completion of the scattering amplitude,
i.e. an automated calculation of the rational part.
The second, far more
difficult, challenge are the phase space integrations of both the
one-loop amplitudes and the bremsstrahlung contributions.

\appendix

\end{document}